\documentclass[pdflatex,sn-nature]{sn-jnl}

\usepackage{graphicx}%
\usepackage{multirow}%
\usepackage{amsmath,amssymb,amsfonts}%
\usepackage{amsthm}%
\usepackage{mathrsfs}%
\usepackage[title]{appendix}%
\usepackage{xcolor}%
\usepackage{textcomp}%
\usepackage{manyfoot}%
\usepackage{booktabs}%
\usepackage{float}
\usepackage{listings}%
\usepackage[ruled,linesnumbered]{algorithm2e}
\usepackage{lineno}

\theoremstyle{thmstyleone}%
\theoremstyle{thmstyletwo}%

\theoremstyle{thmstylethree}%

\unnumbered 

\begin{document}

\title[Article Title]{PhaseT3M: 3D Imaging at 1.6 \AA\ Resolution via Electron Cryo-Tomography with Nonlinear Phase Retrieval}

\author*[1,2]{\fnm{Juhyeok} \sur{Lee}}\email{jhlee0667@lbl.gov}
\author[3]{\fnm{Samuel W.} \sur{Song}}
\author[2]{\fnm{Min Gee} \sur{Cho}}
\author[3,4]{\fnm{Georgios} \sur{Varnavides}}
\author[2]{\fnm{Stephanie M.} \sur{Ribet}}
\author[5]{\fnm{Colin} \sur{Ophus}}
\author*[2,3]{\fnm{Mary C.} \sur{Scott}}\email{mary.scott@berkeley.edu}
\author*[1,6]{\fnm{Michael L.} \sur{Whittaker}}\email{mwhittaker@lbl.gov}

\affil[1]{\orgdiv{Energy Geosciences Division}, \orgname{Lawrence Berkeley National Laboratory}, \orgaddress{\city{Berkeley}, \postcode{94720}, \state{CA}, \country{USA}}}
\affil[2]{\orgdiv{National Center for Electron Microscopy, Molecular Foundry}, \orgname{Lawrence Berkeley National Laboratory}, \orgaddress{\city{Berkeley}, \postcode{94720}, \state{CA}, \country{USA}}}
\affil[3]{\orgdiv{Department of Materials Science and Engineering}, \orgname{University of California Berkeley}, \orgaddress{\city{Berkeley}, \postcode{94720}, \state{CA}, \country{USA}}}
\affil[4]{\orgdiv{Miller Institute for Basic Research in Science}, \orgname{University of California Berkeley}, \orgaddress{\city{Berkeley}, \postcode{94720}, \state{CA}, \country{USA}}}
\affil[5]{\orgdiv{Department of Materials Science and Engineering}, \orgname{Stanford University}, \orgaddress{\city{Stanford}, \postcode{94305}, \state{CA}, \country{USA}}}
\affil[6]{\orgdiv{Materials Sciences Division}, \orgname{Lawrence Berkeley National Laboratory}, \orgaddress{\city{Berkeley}, \postcode{94720}, \state{CA}, \country{USA}}}

\abstract{
Electron cryo-tomography (cryo-ET) enables 3D imaging of complex, radiation-sensitive structures with molecular detail. However, image contrast from the interference of scattered electrons is nonlinear with atomic density and multiple scattering further complicates interpretation. These effects degrade resolution, particularly in conventional reconstruction algorithms, which assume linearity. Particle averaging can reduce such issues but is unsuitable for heterogeneous or dynamic samples ubiquitous in biology, chemistry, and materials sciences. Here, we develop a phase retrieval-based cryo-ET method, PhaseT3M. We experimentally demonstrate its application to an approximately 7 nm Co$_3$O$_4$ nanoparticle on an approximately 30~nm carbon substrate, achieving a maximum resolution of 1.6~\AA, surpassing conventional limits using standard cryo-TEM equipment. PhaseT3M uses a multislice model for multiple scattering and Bayesian optimization for alignment and computational aberration correction, with a positivity constraint to recover `missing wedge’ information. Applied directly to biological particles, it enhances reconstruction quality and reduces artifacts, establishing a new standard for routine 3D imaging with phase contrast.
}


\maketitle

\section{Introduction}\label{Intro}
Transmission electron microscopy (TEM) enables the formation of two-dimensional (2D) structural images from atomic- to micrometer scales, making it a valuable characterization tool for materials of nearly any composition. However, critical structural details such as defects, interfaces, complex architectures, and weak electron signals are hidden by the projection of the three-dimensional (3D) material potential to 2D \cite{Van_Aert_2011, Chen_2013, Renaud_2018, Turk_Baumeister_2020, Lee_2022}. Electron tomography (ET) addresses this limitation by combining images from multiple angles to reconstruct 3D volumes from 2D projections. This approach reveals 3D features that are inaccessible with 2D imaging alone \cite{Midgley_2009, Van_Aert_2011, Scott_2012, Maaz_2015, Ercius_2015, Zhou_Yang_2020, Ilett_2020}, yet are essential for determining material properties.

Recent advances in ET have pushed its resolution to the atomic scale through the use of aberration-corrected annular dark-field (ADF) scanning transmission electron microscopy (STEM) \cite{Rose_1994, Rose_2009, Hetherington_2004, Van_Aert_2011, Scott_2012, Goris_2013, Yang_2017, Zhou_Yang_2020, Lee_2021, Lee_2022}. Despite the advances, ADF-STEM suffers from inherent limitations, including nonlinear contrast due to multiple scattering, a restricted depth of focus, poor sensitivity to light elements, and high electron dose requirements due to slow scanning time. These drawbacks make it less effective for 3D imaging of thick or radiation-sensitive materials \cite{Treacy_Gibson_1993, Hartel_1996, Midgley_Weyland_2003, Xin_Muller_2009, Krivanek_2010, Kirkland_2020, Lee_2023}.

In contrast to scanning techniques, conventional TEM (CTEM) uses a parallel electron beam, allowing for rapid, low-dose imaging of dynamic processes and radiation-sensitive materials \cite{Maaz_2015, Ilett_2020}. CTEM image contrast arises from the interference of scattered electrons from the electrostatic potential. Similar to STEM, stronger electron-atom interactions and increased sample thickness result in more multiple scattering, leading to more nonlinear interference contrast compared to atomic density. Moreover, multiple scattering in samples thicker than approximately twice the mean free electron path \cite{Martynowycz_2021} degrades beam coherence, and therefore limits resolution. Additional contrast modulation is created by lenses in the imaging system. These effects hinder accurate structural reconstruction using conventional tomography techniques that assume linear image contrast \cite{Maaz_2015, Kirkland_2020, Ilett_2020, Lee_2023}, except in special cases \cite{Park_2015, Kim_2020} where electrons scatter only once from a single atom type. The special cases can achieve atomic resolution but require thousands of aberration-corrected high-resolution TEM (HRTEM) images and high electron doses, and are limited to tiny nanoparticles, making them impractical for larger or beam-sensitive samples \cite{Park_2015, Kim_2020}.

Weak electron-atom interactions in organic materials allow their structures to be determined from single-scattering reconstructions \cite{Maaz_2015, Ilett_2020}, but their radiation sensitivity prevents atomic details from being resolved in individual particles, and their image contrast is not strictly linear. Advanced 3D imaging techniques, such as single-particle cryo-electron microscopy (cryo-EM), reconstruct the average 3D atomic positions of particles as small as protons with near-atomic resolution \cite{Kühlbrandt_2014, Cheng_2015, Schur_2016, Nakane_2020, Yip_2020, Liu_2023}, overcoming radiation sensitivity limits. However, these techniques are limited to materials with thousands of nearly identical copies that can be averaged together, which precludes their application to heterogeneous materials.

Interpretable 3D imaging of arbitrary materials requires a physical model to solve the electron-atom scattering problem and instrumental contrast transfer modulation to overcome their limitations, while also being robust to noise arising from sensitive or thick samples. Emerging 3D phase retrieval algorithms based on the multislice method address these limitations by solving the inverse scattering problems and enabling reliable 3D reconstruction of thick samples under low electron doses \cite{Kirkland_2020, Van_den_Broek_Koch_2012, Van_den_Broek_Koch_2013, Ren_2020, Chang_Miao_2020, Whittaker_2022, Lee_2023}. STEM-based 3D phase retrieval has recently made major advances in retrieving 3D phase information with improved resolution and sensitivity to light elements. \cite{Pelz_Ophus_2023, Romanov_Pelz_2024, You_Pelz_2024, Kim_2025}. However, implementing these methods broadly in practice remains challenging, especially for radiation-sensitive materials, due to beam damage from the intense probe, long acquisition times for large areas, and distortions from scan instabilities. In contrast, HRTEM in CTEM enables fast imaging over large fields of view. Combined with fractional imaging and motion correction, it can achieve high resolution without significant loss, even for radiation-sensitive materials \cite{Li_2013, Zheng_2017, Zheng_Wolff_2022}.

Here, we experimentally demonstrate an HRTEM tomography method based on phase retrieval, PhaseT3M, showing that 3D phase-contrast imaging can achieve a maximum resolution of 1.6 \AA\ with a standard cryo-electron microscope without physical aberration correction. We reconstruct a 7 nm Co$_3$O$_4$ nanoparticle supported on 30 nm of carbon using the multislice model to account for multiple scattering and Bayesian optimization to computationally correct image and tilt alignment, contrast transfer, and microscope aberrations. Using the same method, we also reconstruct HIV-1 (human immunodeficiency virus) particles from the publicly available EMPIAR-10164 dataset \cite{Schur_2016}, which represents higher Fourier ring correlation (FRC) values and 20-50\% R-factor improvement compared to conventional reconstruction methods. Our method is applicable to both organic and inorganic materials, vitrified in ice or under vacuum. Autonomous data acquisition and reconstruction scripts are widely available and easily implemented on standard commercial microscopes, making this a powerful scientific tool that spans materials, biological, physical, chemical, and geological sciences.

\section{Results}\label{Results}

\subsection{Tomography Experiment Setup}\label{subsec1}
\begin{figure}[H]\label{fig:tomo_setup}
\centering
\includegraphics[width=0.8\textwidth]{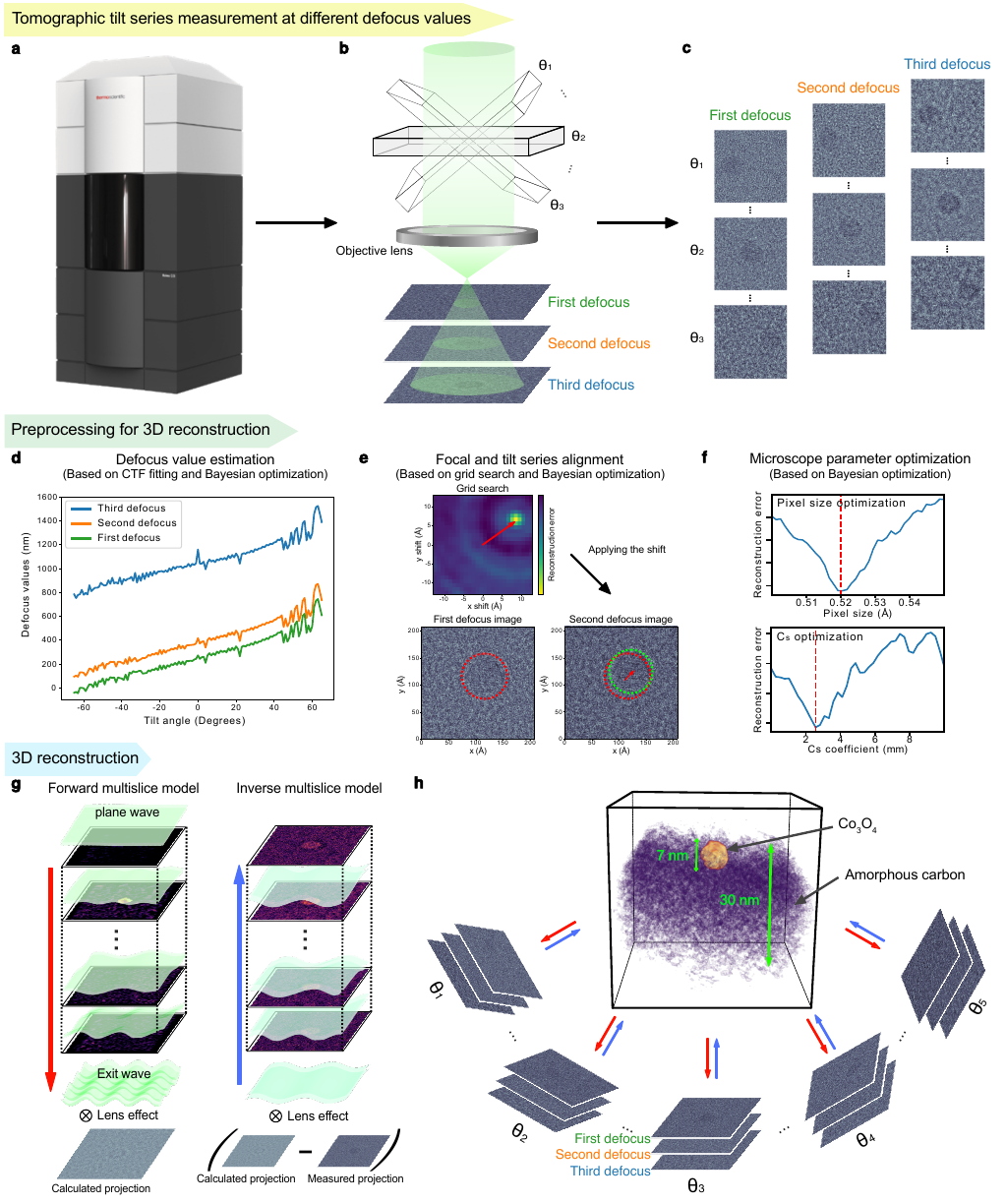}
\caption{
    \textbf{Steps for Experimental Phase Retrieval-based Electron Cryo-Tomography, PhaseT3M.} 
    \textbf{a-c,} Workflow for collecting tomographic tilt-series data at multiple defocus values.
    \textbf{a,} Photograph of a commercial Cryo-TEM setup for measuring HRTEM tomographic data (reproduced with permission from Thermo Fisher Scientific).
    \textbf{b,} Schematic illustration of tomographic tilt-series acquisition at different defocus levels.
    \textbf{c,} Acquisition of HRTEM tilt series images at multiple defocus settings using a standard Cryo-TEM instrument.
    \textbf{d-f,} Preprocessing steps for 3D reconstruction.
    \textbf{d,} Plot illustrating the estimated defocus values at each tilt angle, showing tilt-dependent variations.
    \textbf{e,} Alignment procedure, where the optimal alignment with minimal reconstruction error is identified through a grid search for 2D shifts.
    \textbf{f,} Estimation of microscope parameters via Bayesian optimization to minimize reconstruction errors.
    \textbf{g-h,} Workflow for experimental tomography reconstruction.
    \textbf{g,} Forward and inverse multislice models for solving the scattering problem. At each tilt angle, the multislice forward model calculates simulated HRTEM projections, and the inverse model reconstructs the 3D volume.
    \textbf{h,} Schematic illustration of phase contrast tomography setup. The phase contrast tomography directly reconstructs the 3D potential using multiple tilted focal series of HRTEM images. The 3D volume within the black box represents the reconstructed potential of a Co$_3$O$_4$ nanoparticle embedded in a thick carbon support. Note that, to enhance visualization, distinct color scales are applied to differentiate the Co$_3$O$_4$ nanoparticle from the carbon support. The total 3D reconstruction achieves a voxel size of 2.08~\AA, after applying a binning factor of 4.
    Source data for line plots in this figure are provided as a Source Data file.
}
\end{figure}

\hyperref[fig:tomo_setup]{Figure 1a-c} shows the experiment setup of our 3D phase-contrast tomography, PhaseT3M. We collected a tomographic tilt series of Co$_3$O$_4$ nanoparticles supported by a carbon grid, by acquiring HRTEM images at tilt angles ranging from $-$65° to $+$65° in 1° increments (see \hyperref[fig:Co3O4_tilt_series0]{Supplementary Fig. 1-3}). At each tilt angle, three images were acquired at different defocus values. The electron dose per image was 21.5 $e^{-}$/\AA$^2$, resulting in a total accumulated dose of 8,460 $e^{-}$/\AA$^2$ across the entire series. As shown in \hyperref[fig:tomo_setup]{Fig. 1d-f}, we performed image preprocessing for 3D reconstruction. These steps include motion correction, focal and tilt alignment, estimation of microscopy parameters, and denoising (see \hyperref[Methods]{Methods}). From the preprocessed tilt series, we reconstructed a 3D potential of the sample, which includes an approximately 7 nm diameter Co$_3$O$_4$ nanoparticle and a 30 nm thick amorphous carbon support. To retrieve 3D phase structural information and account for multiple scattering, we applied a forward and inverse multislice approach \cite{Ren_2020, Ren_2021} (see \hyperref[Methods]{Methods}), as illustrated in \hyperref[fig:tomo_setup]{Fig. 1g}. We iteratively reconstructed the 3D potential using the gradient descent method, minimizing the differences between the calculated projections from the estimated 3D potential and the experimentally measured projections.

\subsection{Experimental Tomography via Phase Retrieval}\label{subsec2}
\begin{figure}[H]\label{fig:exp_recon}
\centering
\includegraphics[width=1\textwidth]{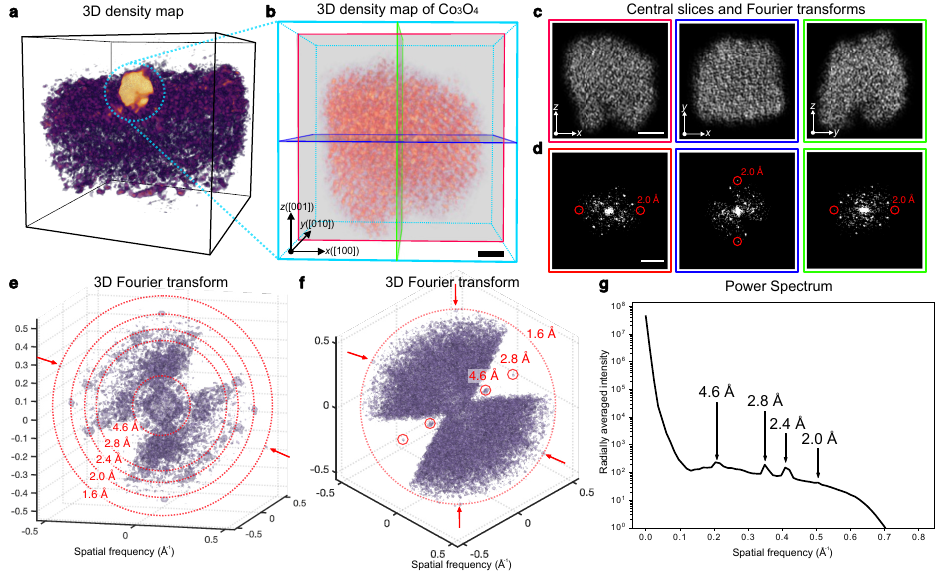}
\caption{
    \textbf{Experimental 3D Reconstruction with Phase Retrieval (PhaseT3M) and Resolution Analysis.} 
    \textbf{a,} Half-sectioned 3D density map showing the internal cross-section of the full 3D reconstruction of a Co$_3$O$_4$ nanoparticle partially embedded in a carbon support.
    \textbf{b,} 3D density map of the Co$_3$O$_4$ nanoparticle after applying a 3D mask to remove the carbon support. Scale bar: 1 nm.
    \textbf{c,} 2-\AA~thick central slices of the 3D reconstruction in (b), with frame colors corresponding to the slice positions in the 3D volume of (b). Image intensities are represented in grayscale. Scale bars: 2 nm.
    \textbf{d,} 2D Fourier transform of the central slice shown in the top panel. Fourier transform intensities are shown on a logarithmic scale in grayscale. Scale bars: 0.4 \AA$^{-1}$.
    \textbf{f,} 3D Fourier transform of the reconstruction in (b), displaying diffraction peaks at up to 1.6 \AA\ resolution.    
    \textbf{g,} 3D Fourier transform of the reconstruction in (b) from a different view, highlighting the missing wedge region. The image shows restored diffraction peaks within the missing wedge and weak diffraction peaks at 1.6 \AA\ resolution. Red arrows indicate the 1.6 \AA\ resolution peaks.
    \textbf{g,} Power spectrum of the 3D reconstruction shown in (b), indicating a resolution limit of 2.0 \AA.
    Source data for line plots in this figure are provided as a Source Data file.
}
\end{figure}

The reconstructed volume in \hyperref[fig:exp_recon]{Fig. 2a} represents a 3D potential of the 7 nm Co$_3$O$_4$ nanoparticle partially embedded in the 30 nm carbon support. The intensity of the reconstructed potential is approximately proportional to the atomic number: lighter elements like carbon and oxygen produce lower potential signal, while heavier elements like cobalt show a higher potential magnitude. As seen in \hyperref[fig:exp_recon]{Fig. 2a}, the Co$_3$O$_4$ nanoparticle and the carbon support are clearly distinguishable. 

To estimate the quality and resolution of the 3D reconstruction, we reconstructed a 3D potential using a cropped tilt series focusing on the Co$_3$O$_4$ nanoparticle with a pixel size of 0.52 \AA\ (see \hyperref[Methods]{Methods}). We then applied a 3D mask to the reconstructed volume in \hyperref[fig:exp_recon]{Fig. 2b} to remove the intensity of the carbon support film. As shown in \hyperref[fig:exp_recon]{Fig. 2c}, the central slices of the 3D reconstructed volume along different axes reveal bright atomic intensities and lattice lines of the nanoparticle. The 2D Fourier transform of each central slice in \hyperref[fig:exp_recon]{Fig. 2d} shows the \{400\} diffraction peak at a resolution of 2 \AA, except for the slice where the missing wedge region overlaps with one of the \{400\} diffraction peaks. The 3D Fourier transform of the 3D reconstructed image in \hyperref[fig:exp_recon]{Fig. 2e} also exhibits diffraction peaks at 2 \AA\ resolution, consistent with the 2D Fourier transform results. \hyperref[fig:exp_recon]{Figure 2f} shows a 3D isosurface image with a lower threshold than that in \hyperref[fig:exp_recon]{Fig. 2e}, presented from a different viewpoint to highlight the missing wedge region. Interestingly, the diffraction peaks of 4.6 \AA\ \{111\} and 2.8 \AA\ \{220\} resolution in the missing wedge region were restored. This improvement is attributed to the application of a physical constraint, positivity (i.e., enforcing nonnegative potential values), which helps suppress missing wedge artifacts and the elongation effect, ultimately aiding in the recovery of diffraction peaks. As shown in \hyperref[fig:positivity]{Supplementary Fig. 4}, only the reconstruction with the positivity constraint exhibits the recovery of diffraction peaks within the missing wedge. Additionally, \{422\} diffraction peaks at 1.6 \AA\ resolution in \hyperref[fig:exp_recon]{Fig. 2e-f} are visible, indicating the high-resolution capability of our reconstruction. This suggests that the maximum resolution of our 3D reconstruction is 1.6 \AA. However, due to the limited tilt range and single, fixed rotation axis, the resolution is anisotropic, varying depending on direction. We also plotted the power spectrum in \hyperref[fig:exp_recon]{Fig. 2g} to cross-check the resolution estimation, and the results are consistent with the previous Fourier transform analysis.

To investigate the effect of multiple scattering correction using the inverse multislice method, we compared slices from volumes reconstructed with the multislice and single-slice phase retrieval approaches. As seen in \hyperref[fig:multi_single]{Supplementary Fig. 5}, multiple scattering correction via the multislice method improves the resolution. Furthermore, \hyperref[fig:error_curve_Co3O4]{Supplementary Fig. 6} shows that the multislice-based method yields a lower reconstruction error compared to the single-slice approach. Additionally, as shown in \hyperref[fig:defocus_variation]{Supplementary Fig. 7}, the multislice method can account for defocus gradients arising from variations in sample height and tilting.

To evaluate the resolution capability of our reconstruction method under different dose conditions, we leveraged our raw image stacks, which consist of 24 subframes per tilt and defocus, allowing us to analyze varying total doses by selecting subsets of these subframes. Reconstructions were performed at total doses of 8460 (24 subframes), 4230 (12 subframes), 705 (2 subframes), 353 (1 subframe), and 118 (one defocus image with one subframe)~$e^{-}$/\AA$^2$. As shown in \hyperref[fig:res_elec_dose]{Supplementary Fig. 8}, the maximum achievable resolution systematically decreases as the total dose decreases. Notably, at the lowest dose of 118~$e^{-}$/\AA$^2$, some diffraction peaks (corresponding to 5.8~\AA~in the 2D Fourier transform and 2.8~\AA~in the 3D Fourier transform) remain visible, but the lattice structure cannot be resolved. These results suggest that while low-dose conditions can retain limited high-resolution information, achieving near-atomic resolution beyond lattice-level resolution requires higher electron doses or averaging across multiple particles to increase the effective signal-to-noise ratio.

\subsection{Simulated Tomography via Phase Retrieval}\label{subsec3}
\begin{figure}[H]\label{fig:simu_tomo}
\centering
\includegraphics[width=1\textwidth]{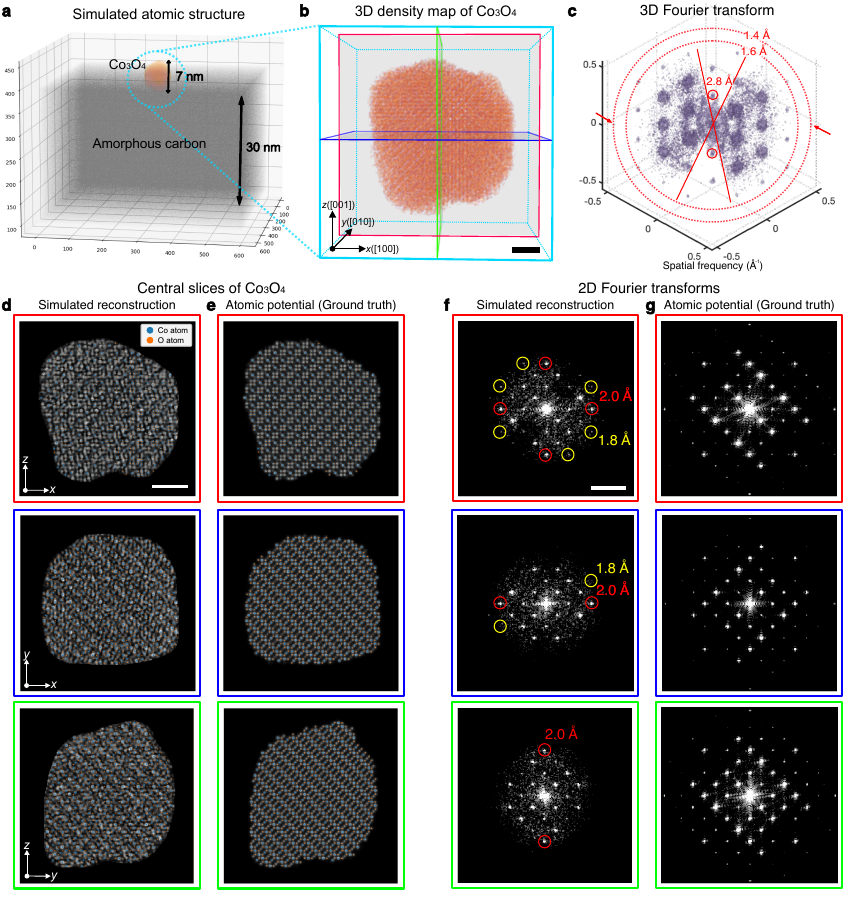}
\caption{
    \textbf{Simulated 3D Reconstruction with Phase Retrieval (PhaseT3M) and Resolution Analysis.} 
    \textbf{a,} 3D atomic models of the Co$_3$O$_4$ nanoparticle and the carbon support. The 3D atomic position of the Co$_3$O$_4$ nanoparticle is based on a perfect crystal, whereas the 3D atomic position of the amorphous carbon support is generated randomly.
    \textbf{b,} 3D density map of the Co$_3$O$_4$ nanoparticle after applying a 3D mask to remove the carbon support. Scale bar: 1 nm.
    \textbf{c,} 3D Fourier transform of the reconstruction in (b), highlighting the missing wedge region. The image reveals restored diffraction peaks within the missing wedge and weak diffraction peaks at 1.4 \AA\ resolution. The red arrows represent the 1.4 \AA\ resolution limit.
    \textbf{d-e,} 2-\AA~thick central slices of the 3D simulated reconstruction (d) and 3D atomic potential volume (e) of the Co$_3$O$_4$ nanoparticle. Each color frame corresponds to the slice color in the 3D reconstruction in (b). Blue and orange dots represent the atomic positions of cobalt and oxygen atoms, respectively. Image intensities are represented in grayscale. Scale bars: 2 nm.
    \textbf{f-g,} 2D Fourier transforms of the central slices from the 3D simulated reconstruction (f) and the 3D atomic potential (g). Red and yellow circles represent the 2.0~\AA\ and 1.8~\AA\ diffraction peaks, respectively. Fourier transform intensities are shown on a logarithmic scale in grayscale. Scale bars: 0.4 \AA$^{-1}$.
}
\end{figure}

To verify the resolution estimated from our experimental reconstruction, we conducted simulations to reproduce our phase retrieval-based tomography experiment. By comparing the experimental and simulated results, we confirmed the validity of our resolution estimation. As shown in \hyperref[fig:simu_tomo]{Fig. 3a}, we first created 3D atomic positions of a Co$_3$O$_4$ nanoparticle and an amorphous carbon membrane (see \hyperref[Methods]{Methods}). Next, we performed TEM simulations to generate the simulated tilt and focal series under the same experimental conditions. We applied Poisson noise to match the electron dose level used in the experiment (see \hyperref[Methods]{Methods}). Finally, we reconstructed a 3D potential from the simulated tilt series using the same reconstruction process as in the experiment (see \hyperref[Methods]{Methods}). We then applied a 3D mask to the reconstructed volume to extract the nanoparticle intensity. Additionally, for using to provide a reference for the reconstruction, we generated 3D atomic potentials from the generated atomic model (see \hyperref[Methods]{Methods}).

Following the same resolution analysis as performed on the experimental results, we evaluated the simulated reconstruction. \hyperref[fig:simu_tomo]{Figure 3b} shows the 3D density map of the reconstructed volume of the Co$_3$O$_4$ nanoparticle, revealing clear lattice fringes. The 3D isosurface plots of the Fourier transform images in \hyperref[fig:simu_tomo]{Fig. 3c} exhibit diffraction peaks at 1.6~\AA\ resolution and recover diffraction peaks within the missing wedge region, consistent with the experimental results. Additionally, faint peaks are observed at 1.4~\AA\ resolution, which is beyond the maximum resolution achieved in the experiment.

The central slices of the 3D volume in \hyperref[fig:simu_tomo]{Fig. 3d} exhibit bright atomic intensities with clear lattice fringes. The simulated atomic structure in \hyperref[fig:simu_tomo]{Fig. 3d,e} aligns well with the intensity maxima in the central slice, with small deviations arising from experimental limitations such as restricted electron dose, elongation from the missing wedge, the thick carbon background, and residual alignment errors. For reference, the ground truth 3D atomic potential in \hyperref[fig:simu_tomo]{Fig. 3e,g} shows perfect agreement with the atomic positions, along with very sharp diffraction peaks up to the Nyquist limit. As seen in \hyperref[fig:simu_tomo]{Fig. 3f}, the \{400\} diffraction peaks at a resolution of 2 \AA\ are visible in the Fourier transform images, except within the missing wedge, similar to the experimental results (\hyperref[fig:exp_recon]{Fig. 2d}). However, simulation results show much sharper diffraction peaks and additional peaks of 1.8~\AA\ beyond the 2.0~\AA\ resolution limit of the experiment. This difference is due to imperfections in alignment, parameter estimation, and noise modeling, as well as limitations of the tomography method, which cannot fully account for all experimental effects, such as inelastic scattering. It is well known that simulation results generally outperform experimental ones because they do not consider all sources of systematic errors and artifacts present in experimental data \cite{Krause_2013}. 
Although the simulation performs slightly better, as expected, the simulated 3D reconstruction, which mimics our experimental conditions, demonstrates results comparable to the experimental data in terms of resolution and diffraction peak retrieval within the missing wedge region.

Additionally, to test the validity of the phase retrieval method, we examined its performance at different electron doses: 8,460~$e^{-}$/\AA$^2$ (used in the experiment), 84,600~$e^{-}$/\AA$^2$, and infinite dose. For simplicity, we used a clean model of Co$_3$O$_4$ nanoparticle without alignment errors and carbon background, including only the missing wedge and electron dose effects. As shown in \hyperref[fig:simul_Co3O4_dose_effect]{Supplementary Fig. 9}, the reconstruction at infinite dose is in agreement with the ground truth, while the reconstruction quality progressively degrades as the dose decreases. Note that applying a low-pass filter of 0.8~\AA$^{-1}$, together with missing wedge effects, hinders the capability to fully resolve the oxygen intensities in the reconstructions. This demonstrates the validity of our method, but also highlights its practical limit under experimental conditions.

\subsection{Comparative analysis of phase retrieval and conventional tomography}\label{subsec4}
\begin{figure}[H]\label{fig:conv_tomo}
\centering
\includegraphics[width=1\textwidth]{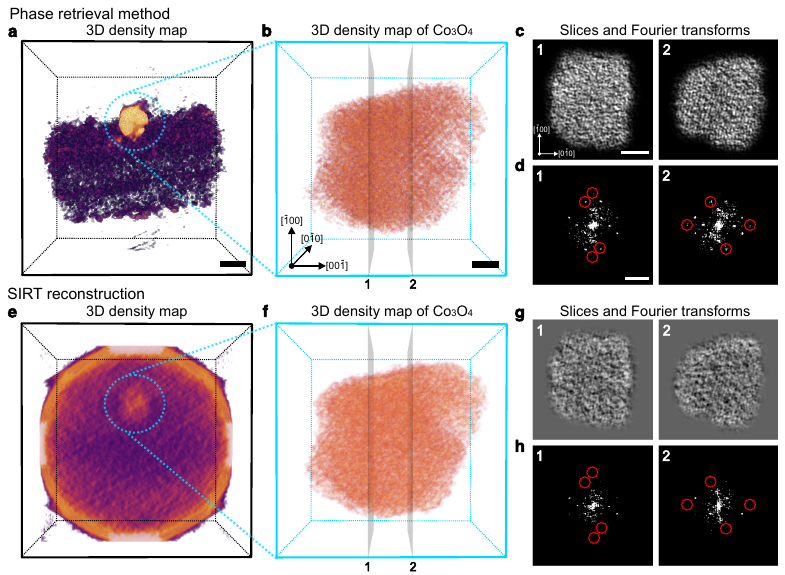}
\caption{
    \textbf{Comparative Analysis of Phase Retrieval (PhaseT3M) and SIRT Reconstructions for the Co$_3$O$_4$ nanoparticle.} 
    \textbf{a, e,} 3D density maps of the Co$_3$O$_4$ nanoparticle and the carbon support reconstructed using phase retrieval (PhaseT3M) (a) and SIRT (e). The 3D volumes were reconstructed with a voxel size of 2.08~\AA. Scale bar: 5 nm.
    \textbf{b, f,} 3D density maps of the Co$_3$O$_4$ nanoparticle reconstructed using phase retrieval (PhaseT3M) (b) and SIRT (f), after applying a 3D mask to remove the carbon support intensity. The 3D volumes were reconstructed with a voxel size of 0.52~\AA. Scale bar: 1 nm.
    \textbf{c, g,} 2~\AA-thick slices extracted from the 3D reconstructions in (b) and (f), respectively. Image intensities are represented in grayscale. Scale bar: 2 nm.
    \textbf{d, h,} 2D Fourier transforms of the slices shown in (c) and (g), respectively. Red circles indicate diffraction peaks present in (d) but absent in (h). Fourier transform intensities are shown on a logarithmic scale in grayscale. Scale bar: 0.4 \AA$^{-1}$.
    The slice numbers in (c–d) and (g–h) correspond to the slice positions indicated in the 3D reconstructions in (b) and (f), respectively.
}
\end{figure}

To compare our approach with conventional tomography, we reconstructed the 3D volume of the Co$_3$O$_4$ nanoparticle and its carbon support using the projection-based tomography algorithm, specifically the simultaneous iterative reconstruction technique (SIRT) (see \hyperref[Methods]{Methods}) \cite{Gilbert_1972}. As shown in \hyperref[fig:conv_tomo]{Fig. 4a,e} and \hyperref[fig:multi_single]{Supplementary Fig. 5a,d}, both the multislice and single-slice phase retrieval methods can retrieve not only the nanoparticle but also the carbon support, whereas conventional tomography fails to reconstruct the carbon support. This suggests that the ability to solve the inverse scattering problem is the critical factor for reconstructing the carbon support, rather than multiple scattering correction. In \hyperref[fig:large_area_rec]{Supplementary Fig. 10c-d}, the carbon intensity is even lower than the vacuum area, and ring-shaped artifacts appear prominently at the boundaries due to nonlinear complex interference contrast, which violates the tomography projection rule.

To further evaluate the performance of the two tomography methods, we compared cropped regions close to the nanoparticle. As shown in \hyperref[fig:conv_tomo]{Fig. 4b,f}, both methods enable the reconstruction of the nanoparticle structures. However, phase retrieval provides higher overall quality and resolution than conventional tomography. This is supported by the Fourier images in \hyperref[fig:conv_tomo]{Fig. 4d,h} and \hyperref[fig:small_area_rec]{Supplementary Fig. 11c,f}, which show higher index diffraction peaks for phase-contrast imaging. Specifically, \hyperref[fig:exp_recon]{Figure 2} demonstrates that our phase retrieval imaging achieves a resolution of 1.6–2.0~\AA, exceeding the typical limits of the conventional tomographic method and the diffraction-limited resolution under our experimental conditions. In comparison, \hyperref[fig:conven_tomo_res]{Supplementary Figure 12c-g} shows that SIRT reconstruction yields a lower resolution of 2.0–2.4~\AA, which is similar to the first zero-crossing point of 2.4~\AA ~indicated by the 2D CTF curve in \hyperref[fig:ctf]{Supplementary Fig. 13}). Additionally, the brighter intensities observed at the nanoparticle boundaries in \hyperref[fig:conv_tomo]{Fig. 4g} and \hyperref[fig:small_area_rec]{Supplementary Fig. 11e} indicate a deficiency in low-frequency information in the SIRT reconstruction. In contrast, the phase retrieval-based reconstruction compensates for these anisotropic intensities by effectively retrieving the missing frequency information through the lens transfer characteristics. This capability can be demonstrated by simulations of the spectral signal-to-noise ratio (SSNR) in 2D focal series reconstruction, presented in \hyperref[fig:SSNR]{Supplementary Fig. 14}. The SSNR indicates that information can be retrieved even at the zero-crossing points of the CTF in raw HRTEM images, where contrast transfer is otherwise absent (see \hyperref[Methods]{Methods}).

\subsection{HIV-1 particles reconstruction}\label{subsec5}
A major advantage of our method is that it can be directly applied to typical HRTEM tomographic tilt data, including open-source datasets such as EMPIAR. Notably, while the previous example used tomographic tilt data acquired with multiple defocus values, our method is also applicable to tilt series collected at a single defocus, enabling the retrieval of the full 3D phase volume.

\begin{figure}[H]\label{fig:bio_tomo}
\centering
\includegraphics[width=1\textwidth]{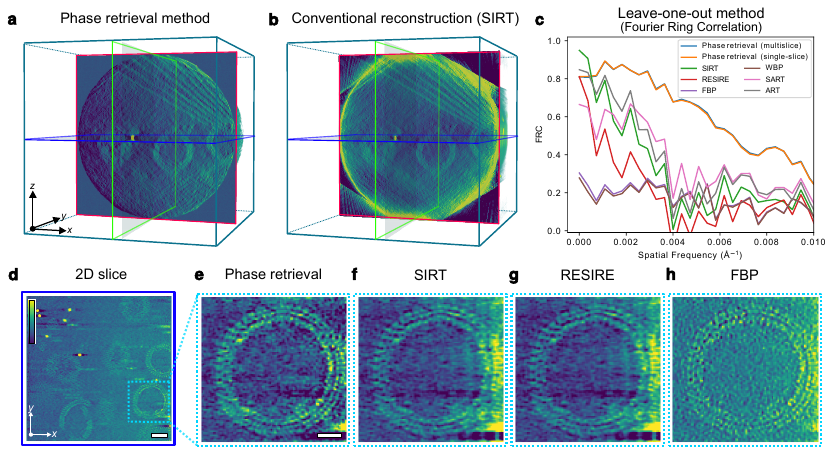}
\caption{
    \textbf{Comparative Analysis of Phase Retrieval (PhaseT3M) and Conventional Reconstructions of HIV-1 Particles (EMPIAR-10164).} 
    \textbf{a-b,} Half-sectioned 3D density maps showing the internal cross-sections of the full 3D reconstructions of HIV-1 particles using our phase retrieval method (PhaseT3M) (a) and the SIRT algorithm (b). Faint, ring-shaped features correspond to HIV-1 particles, while bright, spherical intensities represent gold fiducial markers.
    \textbf{c,} Fourier ring correlation (FRC) between experimental zero-degree HRTEM projection and the corresponding calculated projection from the 3D volume reconstructed from a leave-one-out (zero-degree exclusive) tilt series. FRC curves are compared across different reconstruction methods: multislice phase retrieval (PhaseT3M), single-slice phase retrieval (PhaseT3M), real space iterative reconstruction (RESIRE), simultaneous iterations reconstruction technique (SIRT), filtered back projection (FBP), weighted back projection (WBP), simultaneous algebraic reconstruction technique (SART), and algebraic reconstruction technique (ART).
    \textbf{d,} A 12 nm-thick central slice extracted from the phase retrieval reconstruction, corresponding to the volume shown in (a). The frame color indicates the position of the slice within the 3D volumes. Scale bar: 50 nm.
    \textbf{e-h,} Enlarged views of the boxed region obtained from reconstructions using multislice-phase retrieval (e), SIRT (f), RESIRE (g), and FBP (h) reconstructions, respectively. Scale bars: 10 nm. All color scales represent intensity in the viridis colormap. Source data for line plots in this figure are provided as a Source Data file.
}
\end{figure}

To demonstrate the flexibility of our 3D method, we applied it to a biological sample, HIV-1 particles, using the publicly available EMPIAR-10164 dataset \cite{Schur_2016}. We first prepared a single HIV-1 tomographic tilt-series dataset (\hyperref[fig:HIV_tilt_series]{Supplementary Fig. 15}). We then reconstructed the 3D volumes of HIV-1 particles using the phase retrieval and conventional tomography methods for comparison. \hyperref[fig:bio_tomo]{Figure 5} presents the 3D volumes obtained from our phase retrieval method and the SIRT algorithm \cite{Gilbert_1972}, clearly showing the ring-shaped structures of HIV-1 particles and bright fiducial gold markers. As observed with the Co$_3$O$_4$ nanoparticle reconstructions (\hyperref[fig:conv_tomo]{Fig. 4e}), the conventional tomography reconstructions (\hyperref[fig:bio_tomo]{Fig. 5b} and \hyperref[fig:hiv_projs]{Supplementary Fig. 16c-h}) exhibit noticeable ring-shaped artifacts near the edges because the interference contrast does not strictly satisfy the assumptions of the tomographic projection theorem. In the enlarged views of the reconstructions (\hyperref[fig:bio_tomo]{Fig. 5e–h} and \hyperref[fig:hiv_projs]{Supplementary Fig. 16}), the overall structural features appear similar across different methods. However, conventional tomography reconstructions exhibit pronounced edge artifacts. Notably, iterative-based methods such as SIRT and RESIRE produce smoother intensities due to the suppression of high-frequency noise through iterative updates. In contrast, non-iterative reconstructions, such as FBP and WBP, display sharper intensity features and reduce edge artifacts. However, this apparent improvement can be misleading, as it mainly results from the suppression of low-frequency components by the reconstruction filter, leading to an incomplete recovery of the true structure despite the visually enhanced high-frequency details. This interpretation is supported by our iterative method results using the same reconstruction filter as in FBP and WBP, which produced similar sharpening effects and low-frequency artifact reduction, as shown in \hyperref[fig:ramlak_filter]{Supplementary Fig. 17}.

To evaluate the performance of our approach, we conducted a leave-one-out cross-validation \cite{Cardone_2005} and calculated the R-factor \cite{Yang_2017, Lee_2021}. For the leave-one-out test, we computed FRC curves between the experimental zero-degree projection and the calculated projection from the reconstructed volume, excluding the zero-degree tilt image (see \hyperref[Methods]{Methods}). This approach is particularly effective for assessing biological specimens where the number of tilt projections is limited. As shown in \hyperref[fig:bio_tomo]{Fig. 5c} and \hyperref[fig:average_frc]{Supplementary Fig. 18}, both the multislice and single-slice phase retrieval methods achieve higher correlations compared to conventional tomography algorithms. Notably, the low correlation observed in the low-frequency range of the FRC for FBP and WBP explains their misleading appearance of sharpness, as these methods tend to preserve high-frequency components while suppressing low-frequency information during reconstruction. We also calculated the R-factor between the measured projections and the corresponding forward projections from the reconstructed volumes (see \hyperref[Methods]{Methods}). The multislice phase retrieval method achieved the lowest average R-factor of 3.5\%, compared to 4.7\% for SIRT, 5.8\% for RESIRE, 10.1\% for FBP, 10.1\% for WBP, 5.3\% for SART, and 5.2\% for ART (see \hyperref[fig:rfactor]{Supplementary Fig. 19}). These results indicate that our phase retrieval method demonstrates an enhanced capability to recover missing structural information and provides superior agreement with the experimental projections.

We note that these HIV-1 results are based on a single tilt series, leading to a significantly lower resolution than the previously reported values achieved through subtomogram averaging \cite{Schur_2016}. Furthermore, the tomographic tilt series was acquired with a total electron dose of approximately 100~$e^{-}$/\AA$^2$, which inherently limits the achievable resolution compared to the 1.6~\AA~resolution obtained for the Co$_3$O$_4$ nanoparticle, where two orders of magnitude higher dose was used. It is also important to consider that organic materials like HIV-1 are composed of light elements, which generally exhibit weak electron-atom scattering. This is consistent with our observation that the multislice and single-slice approaches produce nearly identical FRC curves, reconstruction errors, and R-factor, with the multislice results showing a very slight improvement, as shown in \hyperref[fig:rfactor]{Supplementary Fig. 19} and \hyperref[fig:error_curve_EMPIAR]{Supplementary Fig. 20}. Consequently, conventional reconstruction methods, including computationally efficient, non-iterative approaches such as back projection-based methods, can yield reasonably accurate results for such samples, particularly when combined with particle averaging techniques. Nevertheless, the objective of this work is to demonstrate that our physics-based phase retrieval framework can be directly applied to biological datasets and still provide improved resolution and artifact suppression, even from a single tomographic tilt series. Importantly, we anticipate that the benefits of our method will be more pronounced for thicker or more structurally heterogeneous biological specimens, where multiple scattering and interference effects degrade the performance of conventional reconstruction algorithms.

\section{Discussion}\label{Discussion}
In this study, we present PhaseT3M, a 3D phase retrieval-based tomography for high-resolution imaging using multiple tilted HRTEM focal series. By advancing phase retrieval techniques and developing algorithms for sub-pixel alignment and precise parameter optimization, we successfully reconstructed an approximately 7 nm Co$_3$O$_4$ nanoparticle embedded in an approximately 30 nm thick carbon support. This marks the first demonstration of near-atomic resolution 3D phase contrast imaging from a single HRTEM tilt and focal series, overcoming major challenges such as nonlinear artifacts.

Our method introduces several key innovations that contribute to the improved reconstruction: (1) an inverse model that enables accurate 3D potential reconstruction, including weakly scattering components such as the carbon support; (2) a multislice formulation that corrects for nonlinear effects arising from multiple scattering, thereby improving resolution; and (3) a positivity constraint that helps recover missing wedge information, enhancing Fourier completeness and reducing reconstruction artifacts.

Beyond our nanoparticle example, PhaseT3M is broadly applicable to various samples, including both biological samples and nanomaterials. Our approach is compatible with most TEM equipped with direct electron detectors and does not rely on specific hardware configurations. This is particularly valuable for biological imaging, where the availability of extensive open-source datasets enable broad comparative studies. Our method can be directly applied to such datasets, as demonstrated with the HIV-1 tilt series, facilitating high-resolution and artifact-suppressed reconstructions without requiring additional hardware modifications. The method is also compatible with single-particle techniques in cryo-EM and cryo-ET, enabling detailed 3D structural analysis of radiation-sensitive materials.

Although PhaseT3M avoids restrictive approximations and significantly improves accuracy, it currently requires substantial computational resources, approximately 100 times more than conventional methods. However, as GPU technology and computational tools continue to improve, we expect this approach to become increasingly practical and accessible. Additionally, because the phase retrieval reconstructs the atomic potential, the resulting contrast for biological samples composed primarily of light elements is inherently comparable to that of the surrounding vitreous ice, making interpretation more challenging. Although this method increases reconstruction quality, it does not inherently improve contrast for such samples and shows similar contrast levels to other methods. To enhance interpretability, additional post-processing strategies, such as masking and sub-tomogram averaging, may be necessary. In the near future, the ability to extract high-fidelity 3D structural information without simplifying assumptions represents a major step forward, paving the way for new discoveries across materials science, biology, and beyond.

\newpage
\section{Methods}\label{Methods}

\subsection{Sample preparation}\label{Methods:sample_pre}
Co$_3$O$_4$ nanoparticles were synthesized using the sol-gel method \cite{Cho_2024}. In a 100 mL round-bottom flask, a solution of 0.073 g cobalt (II) perchlorate hexahydrate (Aldrich) in 15 mL 1-octanol (Aldrich) was reacted with 2.67 g oleylamine (Acros) and heated to 120 °C in air. As the mixture heated to 80 °C, 0.7 mL of distilled water was introduced, and the reaction was maintained at 120 °C for 6 hours with stirring. The reaction was then cooled to room temperature. To purify the nanoparticles, excess acetone was used initially, followed by a second purification where the nanoparticles were dispersed in hexane and precipitated with ethanol. After each purification, the nanoparticles were collected via centrifugation at $>$200 xg. The supernatant was discarded, and the nanoparticles were redispersed in toluene for drop-casting onto a TEM grid with a carbon substrate.

\subsection{HRTEM data acquisition}\label{Methods:data_acqui}
A tomographic tilt and focal series of a Co$_3$O$_4$ nanoparticle supported on carbon was acquired in electron counting mode using a Titan Krios G3i microscope (Thermo Fisher Scientific) operating at an acceleration voltage of 300 kV, equipped with a Gatan K3 direct electron detector operating in correlated-double sampling mode \cite{Sun_2021}. The data was collected automatically using SerialEM. The tomography experiment was performed under cryogenic conditions without ice encapsulation, utilizing the super-resolution mode of the K3 detector. The super-resolution pixel size was calibrated to be 0.52 \AA. Tilt angles ranged from $-$65° to $+$65° in 1° increments, with a total electron dose of 8,460 $e^{-}$ \AA$^{-2}$. HRTEM images were recorded as a stack of 24 subframes per image. At each tilt angle, a focal series was acquired at initial defocus values of 100 nm, 250 nm, and 900 nm. Note that the final estimated defocus values differed from the initial settings due to changes in defocus resulting from tilting. The estimated defocus values are plotted in \hyperref[fig:tomo_setup]{Fig. 1d}.

A tomographic tilt series of HIV-1 particles (EMPIAR-10164) was obtained from the open-source EMPIAR database \cite{Schur_2016}. We used only the first of the 60 tilt series included in the EMPIAR-10164 dataset for our reconstruction, without using subtomogram averaging. As reported in Ref. \cite{Schur_2016}, the tilt range spans from $-$57° to $+$60° in 3° increments, with a total electron dose of approximately 100 $e^{-}$ \AA$^{-2}$. HRTEM images were recorded as a stack of 8 subframes per image.

\subsection{Preprocessing for 3D reconstruction}\label{Methods:pre}
We applied a series of preprocessing steps to the measured tilt series images, including motion correction, defocus estimation, alignment of focal and tilt series, microscope parameters optimization, and BM3D denoising:

\subsubsection{1) Motion Correction}
To compensate for sample movement during measurement, we employed the motion correction algorithm from Ref. \cite{Li_2013}. Each image in our HRTEM tilt dataset consists of 24 subframes, and for each image, the motion-corrected subframes were summed to produce a single corrected 2D image.

\subsubsection{2) Defocus estimation}
To estimate the defocus values of the HRTEM images, we used CTFFIND4 software \cite{Rohou_2015}, which applies Thon ring fitting. Accurate defocus estimation is crucial for the subsequent focal and tilt alignment steps. Therefore, the fitted defocus values from CTFFIND4 are also later refined during the microscope parameter optimization step, described below.

\subsubsection{3) Focal series alignment}
During the HRTEM acquisition, undesirable image shifts may be introduced in each image along the $x$ and $y$-axes. Correcting these shifts is crucial to achieving accurate, high-resolution reconstructions. We first corrected the alignment within each focal series collected at a given tilt angle, where each focal series consists of three images captured at different defocus values.

To align the images within each focal series, we first performed a grid search over possible 2D shift values ($\Delta x$ and $\Delta y$) between the first and second defocus images. The optimal shift minimizes the reconstruction error of a reduced dataset containing only the first defocus image and the shifted version of the second defocus image. We then applied the same approach to align the first and third defocus images in the same manner, thereby completing the alignment correction for the focal series. This alignment procedure was repeated independently for each tilt angle in our tomographic dataset.

To efficiently determine optimal shifts with sub-pixel accuracy, each grid search was conducted in two stages: an initial coarse search covering a 25 × 25 pixel range with a step size of 2 pixels, followed by a finer search covering a 2 × 2 pixel range with a step size of 0.5 pixels. To reduce the likelihood of being trapped in local minima, we applied a low-pass filter to the defocus images during each search, with a cutoff of 0.3 \AA$^{-1}$ for the coarse search and 0.7 \AA$^{-1}$ for the fine search.

\subsubsection{4) Tilt series alignment}
After correcting the alignment within each focal series, we corrected the alignment between the focal series at different tilts. First, we reconstructed an initial 3D volume using only the focal series at the zero-degree tilt. Next, we performed a grid search over the possible 2D shift values ($\Delta x$ and $\Delta y$) for the focal series at the $+$1° tilt angle. The optimal shift was determined by minimizing the reconstruction error of the $+$1° focal series, using the zero-degree 3D reconstruction as the initial guess.

We then repeated this procedure to obtain the optimal shift for the focal series at the $-$1° tilt angle, this time using a new initial 3D volume reconstructed from both the zero-degree and previously aligned $+$1° tilt series. This alignment process continued sequentially for all remaining tilt angles, sorted by absolute value and alternating between positive and negative tilts (e.g., $+$2°, $-$2°, $+$3°, $-$3°, and so forth). For each tilt angle $\theta$, the optimal shift was identified by minimizing the reconstruction error of the corresponding focal series, with the initial 3D volume guess obtained from the reconstruction of all previously tilt-aligned focal series.
 
Similar to the focal series alignment, we divided each grid search into a coarse search and fine search, with low-pass filter cutoffs of 0.3 \AA$^{-1}$ and 0.7 \AA$^{-1}$, respectively.

\subsubsection{5) Microscope parameters optimization}
We optimized the defocus values, $C_{30}$ value, and pixel size using Bayesian optimization to minimize the reconstruction error (using scikit-optimize python package for Bayesian optimization). Because defocus estimation through Thon ring fitting can be imperfect for thick samples, we refined the three defocus values at each tilt by minimizing the reconstruction error of the focal series at that tilt. These defocus values were further refined using the gradient descent method during the final 3D reconstruction to enhance accuracy. Similarly, we applied Bayesian optimization to determine the optimal $C_{30}$ value and pixel size, which were determined to be 2.3 mm and 0.52 \AA, respectively. The estimated defocus and $C_{30}$ values from this step were used as initial parameters for the 3D reconstruction. During phase retrieval-based reconstruction, these values were further refined using a gradient descent method.

\subsubsection{6) Image denoising}
Before 3D reconstruction, we applied a denoising procedure to each HRTEM image as follows. The HRTEM experimental images are primarily affected by Poisson noise, which can be described as:
\begin{equation} \label{eq10}
    Y = \alpha P (n),
\end{equation}
where $Y$ is the intensity at each pixel, $\alpha$ is the gain parameter, and $P(n)$ is the Poisson distribution of $n$ electrons. We estimated the gain parameter $\alpha$ through statistical analysis of the 24 subframes across each image, following previous works \cite{Yang_2017, Lee_2021, Lee_2022}. The images were denoised by the block-matching and 3D filtering (BM3D) algorithm \cite{Dabov2007}. To meet the BM3D algorithm’s requirement for Gaussian noise, we applied the Anscombe transformation \cite{Makitalo_Foi_2013} to convert the Poisson noise into pure Gaussian noise. After denoising with BM3D, we applied the inverse Anscombe transformation.

\subsection{PhaseT3M, Phase Retrieval algorithm}\label{Methods:recon_method}
By using the HRTEM multislice method as a forward model, we formulate an inverse problem to retrieve the 3D electrostatic potential $V$ from 2D HRTEM images acquired at various tilt angles and focal planes, as illustrated in \hyperref[fig:tomo_setup]{Fig. 1b-c}. This inverse problem is posed as an optimization task that minimizes the error function $\mathcal{E}^2$, which quantifies the discrepancy between the experimentally measured HRTEM images and the calculated HRTEM images from the 3D potential $V$:
\begin{equation} \label{eq1}
    V = \arg\min_{V} \sum_{\theta, \Delta f} \mathcal{E}_{\theta, \Delta f}^2
    = \arg\min_{V} \sum_{\theta, \Delta f} {\Vert \sqrt{ \hat{I}_{\theta, \Delta f}(\vec{r}; V)} - \sqrt{I_{\theta, \Delta f}(\vec{r})}} \Vert_2^2
\end{equation}
, where $\Vert \cdot \Vert_2$ represents the L$_2$ norm, and ${\hat{I}}_{\theta, \Delta f}(\vec{r}; V)$ and $I_{\theta, \Delta f}(\vec{r})$ are the calculated and measured HRTEM image at each tilt angle $\theta$ and each defocus value $\Delta f$, respectively.

Our reconstruction algorithm for solving the inverse problem has three main components: 1) the forward multislice model to calculate HRTEM images, 2) the inverse multislice model to retrieve a 3D electrostatic potential $V$, and 3) regularization to account for experimental noise and enhance the quality of the reconstructed volume. These three steps are iteratively applied across all tilt angles until the error converges sufficiently, yielding the final 3D electrostatic potential $V$. For further detail, see the pseudocode presented in the Supplementary Information.

\subsubsection{1) Forward multislice model}
The forward multislice model uses the current estimate of the 3D electrostatic potential $V$ to calculate HRTEM images, denoted as ${\hat{I}}$. To describe the scattering interactions between the electron beam and the sample, we begin by considering the Schr\"{o}dinger equation for fast electrons within the framework of the forward model \cite{Kirkland_2020}:
\begin{equation} \label{eq2}
\frac{\partial \psi(x, y, z)}{\partial z} = \frac{i \lambda}{4 \pi} \nabla_{xy}^2 \psi(x, y, z) + i \sigma V(x, y, z) \psi(x, y, z),
\end{equation}
where $\psi(x,y,z)$ is the electron wave function, $\lambda$ is the relativistic electron wavelength, $\sigma$ is the relativistic interaction constant. To efficiently solve the Schr\"{o}dinger equation for fast electrons, we use the multislice method, which provides an effective numerical approximation for the beam-sample interaction \cite{Cowley_Moodie_1957} by discretizing the full 3D potential into a stack of two-dimensional (2D) projected potentials. This approach facilitates the calculation of simulated HRTEM images by sequentially propagating the electron wavefunction through each potential slice, accounting for both phase shifts and scattering events.

Initially, the 3D electrostatic potential $V$ is discretized into a series of thin slices, each with a thickness $\Delta z$, along the direction of the electron beam. For each slice $m$, the potential is projected by integrating $V$ over the slice thickness, yielding a 2D projected potential:
\begin{equation} \label{eq3}
     V_m^\textrm{2D}\left(x,y\right) = \int_{z_m}^{z_m+\Delta z}{V(x,y,z)\ dz} 
\end{equation}
where $m = 1$ to $N_z$ represents the slice index.

Next, the evolution of the electron wave function is computed by sequentially applying the free-space propagation function and transmission function for each slice of the material. Specifically, $\psi_{m+1}$, an electron wave function after transmitting through the $m$-th slice, can be expressed using $\psi_m$ as
\begin{equation} \label{eq4}
    \psi_{m+1,{\vec{r}}_p}\left(\vec{r}\right) = \mathcal{F}^{-1}[P\left(\vec{q}\right)\mathcal{F}[t_m\left(\vec{r}\right)\psi_{m,{\vec{r}}_p}(\vec{r})]],
\end{equation}
where $\mathcal{F}$ and $\mathcal{F}^{-1}$ represent the forward and inverse Fourier transform, respectively. The free-space propagation function $P(\vec{q})$ can be written as $P\left(\vec{q}\right)=\textrm{exp}{\left(-i\pi\Delta z (q_x^2+q_y^2)\right)}$, with $\vec{q} = \left(q_x, q_y\right)$ representing the 2D spatial frequency components in the $x$ and $y$ directions, respectively. The transmission function $t_m\left(\vec{r}\right)$ is given as $t_m\left(\vec{r}\right) = \textrm{exp}{\left(i\sigma V_m^\textrm{2D}(x,y)\right)}$. While the slice thickness $\Delta z$ can be allowed to vary between slices to accommodate different material properties or experimental conditions, we opted to use a uniform $\Delta z$ thickness across all slices in our calculation to simplify the computational model.

Finally, the incident parallel electron wave function, $\psi_{0}$, is initialized to unity in the HRTEM forward model. The exit wave function, $\psi_\textrm{exit}$ (also denoted as $\psi_{N_z+1}$), represents the electron wave after transmission through the sample. To obtain  $\psi_\textrm{exit}$, we recursively apply Eq.~\ref{eq4} starting from the incident wave function $\psi_{0}$ and propagate it through each of the $N_z$ slices of the sample, where $N_z$ is the total number of slices. To incorporate defocus effects and other lens aberrations, the final wave function is expressed as:
\begin{equation} \label{eq5}
	\psi_{\textrm{final}, \theta, \Delta f}\left(\vec{r}\right) = \mathcal{F}^{-1}[\psi_{\textrm{exit}, \theta}\left(\vec{q}\right) \exp{(-i \chi \left(\vec{q}\right) )}],
\end{equation}
where $\chi(\vec{q})$ is the aberration function in Krivanek notation \cite{Krivanek_1999, Chang_Miao_2020}: 
\begin{equation} \label{eq6}
    \chi(q, \phi) = \frac{2\pi}{\lambda} \sum_{n,n'} \frac{(\lambda q)^{n+1}}{n+1} \left( C_{nn',a} \cos{(n'\phi)} + C_{nn',b} \sin{(n'\phi)} \right),
\end{equation}
with non-negative integers $n$ and $n'$ defined as follows:
\begin{align*}
    n &= 0, 1, 2, 3, \dots, \\
    n' &=
    \begin{cases}
        0, 2, 4, \dots, (n+1), & \text{if } n \text{ is odd}, \\
        1, 3, 5, \dots, (n+1), & \text{if } n \text{ is even}.
    \end{cases}
\end{align*}
Here, $q = \sqrt{q_x^2+q_y^2}$ is the radial spatial frequency and $\phi = \arctan(q_y / q_x)$ is the azimuthal angle. The coefficients $C_{nn',(a,b)}$ represent aberration coefficients for different types and symmetries of aberrations. For example, $C_{10}$ is equal to the negative of the defocus $-\Delta f$; $C_{01,(a,b)}$ describes the image shift; $C_{12,(a,b)}$ describes twofold astigmatism; $C_{21,(a,b)}$ describes comma; and $C_{30}$ describes third-order spherical aberration. The calculated HRTEM image $\hat{I}_{\theta, \Delta f}$ at each tilt angle and defocus value can be obtained by taking the absolute square value of $\psi_{\textrm{final}, \theta, \Delta f}$. 

\subsubsection{2) Inverse multislice model}
The inverse multislice model iteratively updates the 3D electrostatic potential $V$ to minimize the error function $\mathcal{E}^2$ as defined in Eq.~\ref{eq1} via the gradient descent method. The analytical expression for the derivative of the error function with respect to the m-th 2D projected potential $V_m$ of the 3D potential $V$ is given by:
\begin{equation} \label{eq7}
    \begin{split}
        \nabla_{V_m} \mathcal{E}_{\theta}^2 &= \operatorname{Re} \left( -2i\sigma \Delta z ~ t_m^* \psi_m^* \mathcal{F}^{-1} P^* \mathcal{F} \dots t_{N_z - 1}^* \mathcal{F}^{-1} P^* \mathcal{F} \right. t_{N_z}^* \mathcal{F}^{-1} P^* \\
        &\quad \times \sum_{\Delta f} \exp\left(i \chi{(\Delta f)}\right) \left. \mathcal{F} \left( \psi_{\text{final}, \theta, \Delta f} - \sqrt{I_{\theta, \Delta f}}  \frac{\psi_{\text{final}, \theta, \Delta f}}{\left| \psi_{\text{final}, \theta, \Delta f} \right|}  \right) \right),
    \end{split}
\end{equation}
where the multiplications and divisions in Eq.~\ref{eq7} represent element-wise (point-wise) operations. To simultaneously refine the image shift $C_{01}$, defocus values $C_{10}$ and higher-order aberration coefficients $C_{nn'}$, we derive the gradient of the error function with respect to each aberration coefficient $C_{nn'}$ for each image $I_{\theta, \Delta f}$ as follows:
\begin{equation} \label{eq8}
    \begin{split}
        \nabla_{C_{nn',x}^{(\theta, \Delta f)}} \mathcal{E}_{\theta}^2 &= \sum_{\vec{q}} \operatorname{Re} \left( \frac{4\pi i}{\lambda} \frac{(\lambda q)^{n+1}}{n+1} \right.
        \times \left\{ \begin{array}{ll}
            \cos{(n'\phi)}, & \text{if } x = a \\
            \sin{(n'\phi)}, & \text{if } x = b 
        \end{array} \right\} \\
        &\quad \times(\mathcal{F}\psi_{\text{exit}, \theta})^* \exp\left(i \chi{(\Delta f)}\right) \left. \mathcal{F} \left( \psi_{\text{final}, \theta, \Delta f} - \sqrt{I_{\theta, \Delta f}}  \frac{\psi_{\text{final}, \theta, \Delta f}}{\left| \psi_{\text{final}, \theta, \Delta f} \right|}  \right) \right),
    \end{split}
\end{equation}
where $q = \sqrt{q_x^2+q_y^2}$ is the radial spatial frequency and $\phi = \arctan(q_y / q_x)$ is the azimuthal angle. Using the derived gradient formula, the target quantity $U$ such as $V$ and $C_{nn'}$ is iteratively updated as:
\begin{equation} \label{eq9}
    U = U - \alpha_{U} \nabla_{U} \mathcal{E}_{\theta}^2,
\end{equation}
where $\alpha_{U}$ is the step size controlling the strength of gradient descent. We use Eq.~\ref{eq9} to update the 3D potential $V$ by optimizing image shift, defocus values, and higher-order aberration coefficients across the entire tilt series.

\subsubsection{3) Regularization}
Regularization is incorporated to mitigate experimental noise and enhance the quality of the reconstructed volume. Based on the physical assumption that the 3D electrostatic potential is non-negative and real \cite{Kirkland_2020}, we apply a positivity constraint, setting any negative potential values to zero at each iteration. Additionally, to suppress high-frequency noise generated during iterative reconstruction, a 3D low-pass filter is applied to the reconstructed volume at each iteration.

\subsection{3D phase retrieval-based reconstruction of a Co$_3$O$_4$ nanoparticle}\label{Methods:recon}
Using the 3D phase retrieval algorithm described above, we reconstructed the 3D potential of a Co$_3$O$_4$ nanoparticle supported on carbon. We used $N_z$ = 600 slices for the multislice calculation, 100 iterations, a positivity constraint, and a 3D low-pass filter cutoff of 0.8 \AA$^{-1}$ to reduce high-frequency noise. Note that in \hyperref[fig:positivity]{Supplementary Fig. 4b}, the only difference is that the positivity constraint was not applied. In \hyperref[fig:multi_single]{Supplementary Fig. 5d-f}, a single-slice calculation was used with $N_z = 1$. The resulting size of the reconstructed 3D volume is $600 \times 600 \times 600$ voxels, with a voxel size of 0.52 \AA. The error curves for both multislice and single-slice reconstructions converge, as shown in \hyperref[fig:error_curve_Co3O4]{Supplementary Fig. 6}. In addition to reconstructing the 3D potential, we further refined the image shifts, defocus values, and high-order aberrations (up to $C_{32}$) during the process. For the larger field of view reconstruction, as shown in \hyperref[fig:tomo_setup]{Fig. 1h}, \hyperref[fig:exp_recon]{Fig. 2a}, and \hyperref[fig:conv_tomo]{Fig. 4a}, the 3D volume was reconstructed with pixel binning of 4, yielding a voxel size of 2.08 \AA~and a volume size of $225 \times 225 \times 225$ voxels.

The iterative phase-retrieval problem is inherently non-convex, and convergence to a global minimum cannot be guaranteed in theory.  To mitigate the risk of becoming trapped in local minima, we implemented several practical strategies to ensure reliable convergence: (1) validation on simulated datasets with known ground truth, (2) monitoring smooth error reduction over iterations, (3) random shuffling of tilt projections to avoid bias, and (4) using physics-based constraints and the Adam optimizer to enhance stability. Together, these strategies provide strong empirical evidence that our reconstructions converge reliably without being trapped in poor local minima.

\subsection{Simulation to reproduce the phase retrieval-based tomography experiment}\label{Methods:simul_recon}
To reproduce the phase retrieval-based tomography experiment, we first created a 3D atomic model of a Co$_3$O$_4$ nanoparticle on a carbon support. The Co$_3$O$_4$ nanoparticle was modeled as a spherical, 8 nm-diameter crystal with a perfect Co$_3$O$_4$ structure. For the carbon support, we generated an amorphous structure by randomly placing atoms within a 60 x 60 x 30 nm$^3$ volume, maintaining an atomic density of 100 atoms per nm$^3$. As shown in \hyperref[fig:simu_tomo]{Fig. 3a}, the nanoparticle was placed on the amorphous carbon support, positioned halfway embedded in the carbon structure. Atomic positions of carbon within the overlapping region of the nanoparticle and the support were removed.

Using the generated atomic structure, we simulated a tilt and focal series using the multislice-based HRTEM simulation implemented in the abTEM python package \cite{Madsen_Susi_2021}. We used tilt angles ranging from $-$65° to $+$65° in 1° increments, and we applied the same refined defocus values we obtained from our experimental dataset. To mimic experimental conditions, Poisson noise was applied to the images based on our experimental electron dose, and random image shifts were introduced following a Gaussian distribution with a mean of 0 pixels and a standard deviation of 0.5 pixels.

Using this generated tilt and focal series, we reconstructed the simulated 3D potential. The reconstruction parameters matched those used in the experiment, including 600 slices for the multislice calculation, 100 iterations, and a 3D low-pass filter cutoff of 0.8 \AA$^{-1}$. The resulting full volume size was $600 \times 600 \times 600$ voxels, with a voxel size of 0.52 \AA. A 3D mask was then applied to the reconstructed volume to isolate only the Co$_3$O$_4$ nanoparticle intensity, as shown in \hyperref[fig:simu_tomo]{Fig. 3b}.

As a simulated reconstruction reference, we generated a very sharp 3D potential from the generated atomic model using abTEM \cite{Madsen_Susi_2021}. We then applied Gaussian smoothing to this 3D potential with a standard deviation of 0.65 pixels.

\subsection{SIRT reconstruction of a \texorpdfstring{Co$_3$O$_4$}{Co3O4} nanoparticle}\label{Methods:conven_recon}
We reconstructed a 3D volume from the tilt and focal series after the preprocessing step using the SIRT algorithm \cite{Gilbert_1972} with 200 iterations. To suppress high-frequency noise, a 3D low-pass filter with a cutoff frequency of 0.8 \AA$^{-1}$ was applied. The final reconstruction volume size was $600 \times 600 \times 600$ voxels, with a voxel size of 0.52 \AA. For a larger field of view, as shown in \hyperref[fig:exp_recon]{Fig. 2a}, \hyperref[fig:conv_tomo]{Fig. 4a}, and \hyperref[fig:conven_tomo_res]{Supplementary Fig. 12a}, the 3D volume was reconstructed with pixel binning of 4, resulting in a voxel size of 2.08 \AA~and a volume size of $225 \times 225 \times 225$ voxels.

\subsection{Spectral signal-to-noise ratio (SSNR) calculation}\label{Methods:ssnr}
To calculate the SSNR, we first generated a 2D white noise potential with a size of 100 $\times$ 100 pixels. Using the abTEM Python package \cite{Madsen_Susi_2021}, we simulated HRTEM focal series from the white noise potential with defocus values of 100 nm, 250 nm, and 900 nm. Poisson noise, corresponding to electron dose, was applied to the simulated images, and 50 ensemble focal series were generated. We then reconstructed 50 potentials from these noisy focal series using PhaseT3M. Finally, the SSNR was computed from the reconstructed potentials following the method described in Ref. \cite{Unser_1987}.

\subsection{HIV-1 reconstruction}\label{Methods:bio_recon}
A tomographic tilt series of HIV-1 particles (EMPIAR-10164), as described in the HRTEM data acquisition section of Methods, was used for reconstruction. Motion correction was performed using the algorithm from Ref. \cite{Li_2013}, applied to the 8 subframes of each image in the tilt series. The motion-corrected subframes were then summed to generate a single corrected image per tilt. Defocus values were estimated using CTFFIND4 (version 4.1.14)\cite{Rohou_2015}. Tilt alignment was carried out in IMOD (version 5.1)\cite{Kremer_1996, Mastronarde_2017}, using a cross-correlation-based coarse alignment followed by fiducial marker-based fine alignment. Before reconstruction, we applied binning by a factor of 9, taking the pixel size from 1.35 \AA\ to 12.15 \AA. Various reconstruction methods were independently applied to the binned tilt series data, including phase retrieval, simultaneous iterative reconstruction technique (SIRT) \cite{Gilbert_1972}, real space iterative reconstruction (RESIRE) \cite{Pham_2023}, filtered back projection (FBP), weighted back projection (WBP) \cite{Radermacher_1992}, simultaneous algebraic reconstruction technique (SART) \cite{Andersen_1984}, and algebraic reconstruction technique (ART) \cite{Gordon_1970}. For the reconstructions, we used the tomopy package (version 1.15.2) for SIRT and ART, the ASTRA toolbox (version 2.3.1) for FBP, WBP, and SART, and the RESIRE MATLAB code \cite{Pham_2023} for the RESIRE reconstruction. The error curves for both multislice and single-slice phase retrieval reconstructions converge, as shown in \hyperref[fig:error_curve_EMPIAR]{Supplementary Fig. 20}. The final reconstructed volume size was $360 \times 360 \times 360$ voxels, with a voxel size of 12.15 \AA. We performed 100 iterations for our phase retrieval method and 500 iterations for the SIRT, RESIRE, and SART reconstructions. For the ART reconstruction, 20 iterations were used, yielding the clearest image. Note that all reconstructions of HIV-1 particles were performed without applying a positivity constraint, as variations in vitreous ice thickness can lead to regions of negative potential.

\subsection{Leave-one-out method}\label{Methods:FRC}
To avoid overfitting and to objectively evaluate the reconstruction quality, we adopted a leave-one-out strategy. In this approach, each projection is systematically excluded from the reconstruction process, and a forward projection is generated from the resulting volume for comparison with the omitted experimental image. To calculate the leave-one-out Fourier ring correlation (FRC), we first prepared one measured projection and the corresponding calculated linear projection generated from the 3D volume reconstructed using all tilt projections except for the measured projection, ensuring an unbiased comparison. To minimize Fourier artifacts, the prepared projections were padded using the calculated mean intensity as a constant value to match the background. The FRC was then computed as:
\begin{equation}
FRC(r) = \frac{
\sum\limits_{r_i \in r} F_1(r_i) \cdot F_2(r_i)^*
}{
\sqrt{
\sum\limits_{r_i \in r} |F_1(r_i)|^2 \cdot \sum\limits_{r_i \in r} |F_2(r_i)|^2
}
},
\end{equation}
where $F_1$ and $F_2$ are the Fourier transforms of the two 2D images, and $F_2^*$ is the complex conjugate of $F_2$. The index $r_i$ refers to an individual pixel at radius $r$ in Fourier space. The FRCs shown in \hyperref[fig:bio_tomo]{Fig. 5c} and \hyperref[fig:average_frc]{Supplementary Fig. 18} were calculated using the zero-degree projection.

\subsection{R-factor calculation}\label{Methods:Rfactor}
The R-factor is a metric used to quantify the difference between two images. To evaluate the similarity between the measured projection and the calculated projection from the reconstructed volume, we computed the R-factor as:
\begin{equation}
R = \frac{
    \sum_{i} \left| I^{\text{meas}}_i - I^{\text{calc}}_i \right|
}{
    \sum_{i} \left| I^{\text{meas}}_i \right|
},
\end{equation}
where $I^{\text{meas}}_i$ and $I^{\text{calc}}_i$ represent the measured and calculated intensities at pixel $i$, respectively. The index $i$ runs over all pixels in real space. Note that calculated projections $I^{\text{calc}}_i$ were generated from a forward multislice or single-slice model for PhaseT3M, and a linear projection model for all other conventional reconstruction methods. \hyperref[fig:rfactor]{Supplementary Fig. 19} presents the R-factors calculated for all tilt angles across the different reconstruction methods.

\newpage
\section{Figures}\label{Figs}

\backmatter
\newpage
\section*{Data availability}
Source data for line plots are provided with this paper. Our experimental tilt data and tomographic reconstructions supporting the findings of this study are publicly accessible in the Zenodo repository at \href{https://doi.org/10.5281/zenodo.17336678}{https://doi.org/10.5281/zenodo.17336678}.

\section*{Code availability}
The PhaseT3M source code is available in the GitHub repository at \href{https://github.com/PhaseT3M/phaset3m}{https://github.com/PhaseT3M/phaset3m}.

\newpage
\section*{Acknowledgements}
The work of J.L. and M.L.W. was supported by the U.S. Department of Energy, Office of Science, Office of Basic Energy Sciences, Early Career Research Program and Geosciences programs under Award No. DE-AC02-05-CH11231. 
S.W.S. is supported by the National Science Foundation Graduate Research Fellowship Program under Grant No. DGE 2146752. Any opinions, findings, conclusions or recommendations expressed in this material are those of the authors and do not necessarily reflect the views of the National Science Foundation.
M.C.S. and M.G.C. are funded by the U.S. Department of Energy in the program “4D Camera Distillery: From Massive Electron Microscopy Scattering Data to Useful Information with AI/ML".
This work at the Molecular Foundry was supported by the Office of Science, Office of Basic Energy Sciences, of the U.S. Department of Energy under Contract No. DE-AC02-05-CH11231.
This work used resources of the National Energy Research Scientific Computing Center (NERSC), a Department of Energy User Facility using NERSC award BES-ERCAP 0032753.
This work was carried out using shared experimental facilities at the California Institute for Quantitative Biosciences at the University of California, Berkeley.
We thank Dr. Dan Toso and Dr. Ravi Thakkar for their support with acquiring the experimental cryo-ET data. 
We declare that the authors utilized ChatGPT (https://chat.openai.com/chat) for language editing purposes only, and the original manuscript texts were all written by human authors, not by artificial intelligence.

\section*{Author contribution}
J.L., M.C.S., and M.L.W. conceived the idea and directed the study. M.G.C. synthesized the cobalt oxide nanoparticle. J.L., S.W.S., and M.L.W. designed and performed the tomography experiment. J.L., S.W.S., G.V., S.R., C.O., and M.L.W. developed and enhanced the reconstruction algorithm. J.L. conducted experimental and simulational data analysis. J.L., M.C.S., and M.L.W. wrote the manuscript. All authors commented on the manuscript.

\section*{Competing Interests}
The authors declare no competing interests.

\newpage

\newpage
\section{Supplementary information}\label{SI}
\subsection{Supplementary Figures} 
\setcounter{figure}{0} 
\renewcommand{\thefigure}{S\arabic{figure}} 

\begin{figure}[H]\label{fig:Co3O4_tilt_series0}
\centering
\includegraphics[width=0.9\textwidth]{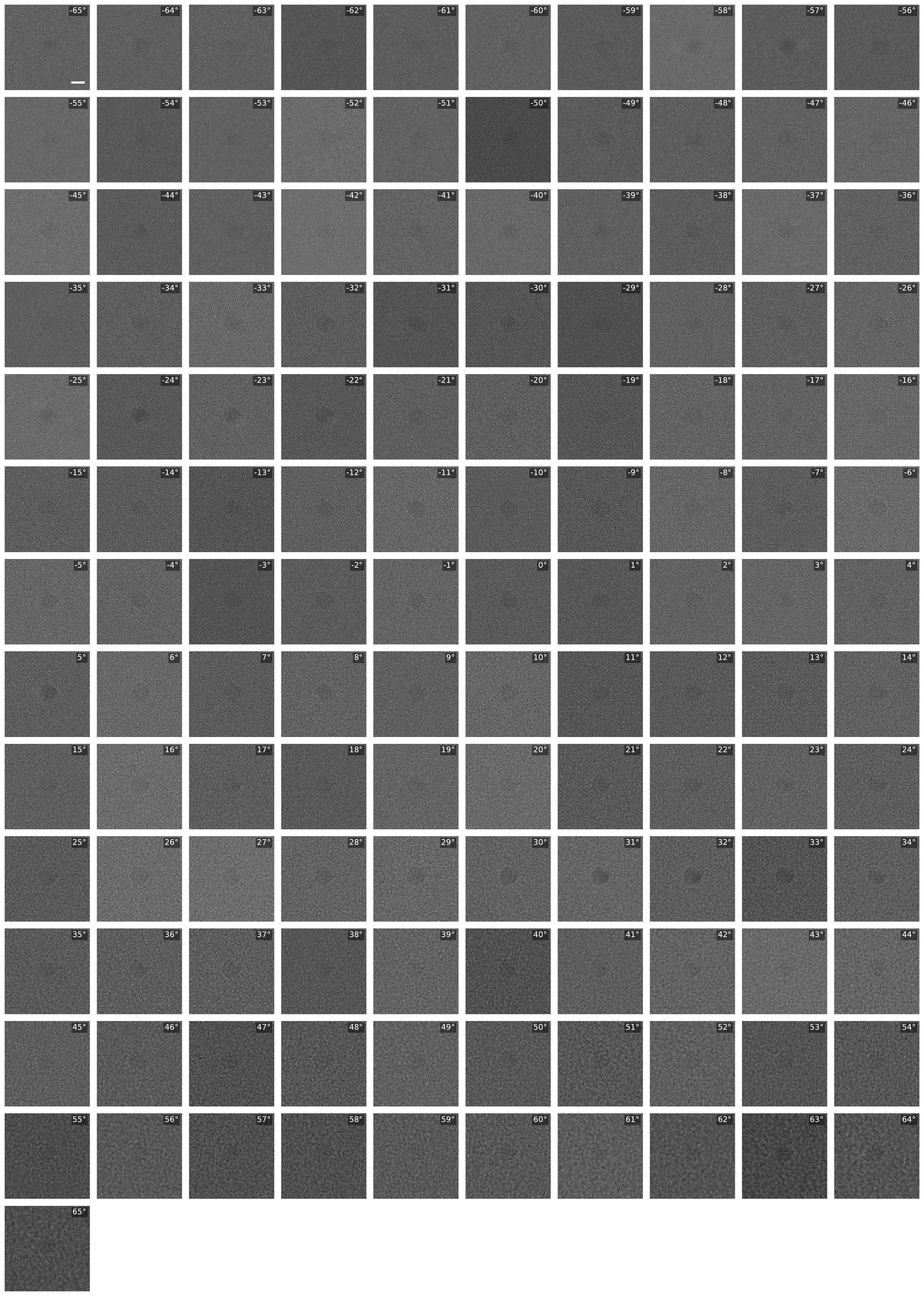}
\caption{
    \textbf{Experimental HRTEM tomographic tilt-series of a Co$_3$O$_4$ nanoparticle (first defocus).} 
    Each image represents a motion-corrected sum of frames. Image intensities are represented in grayscale. Scale bar: 5 nm.
}
\end{figure}

\begin{figure}[H]\label{fig:Co3O4_tilt_series1}
\centering
\includegraphics[width=0.9\textwidth]{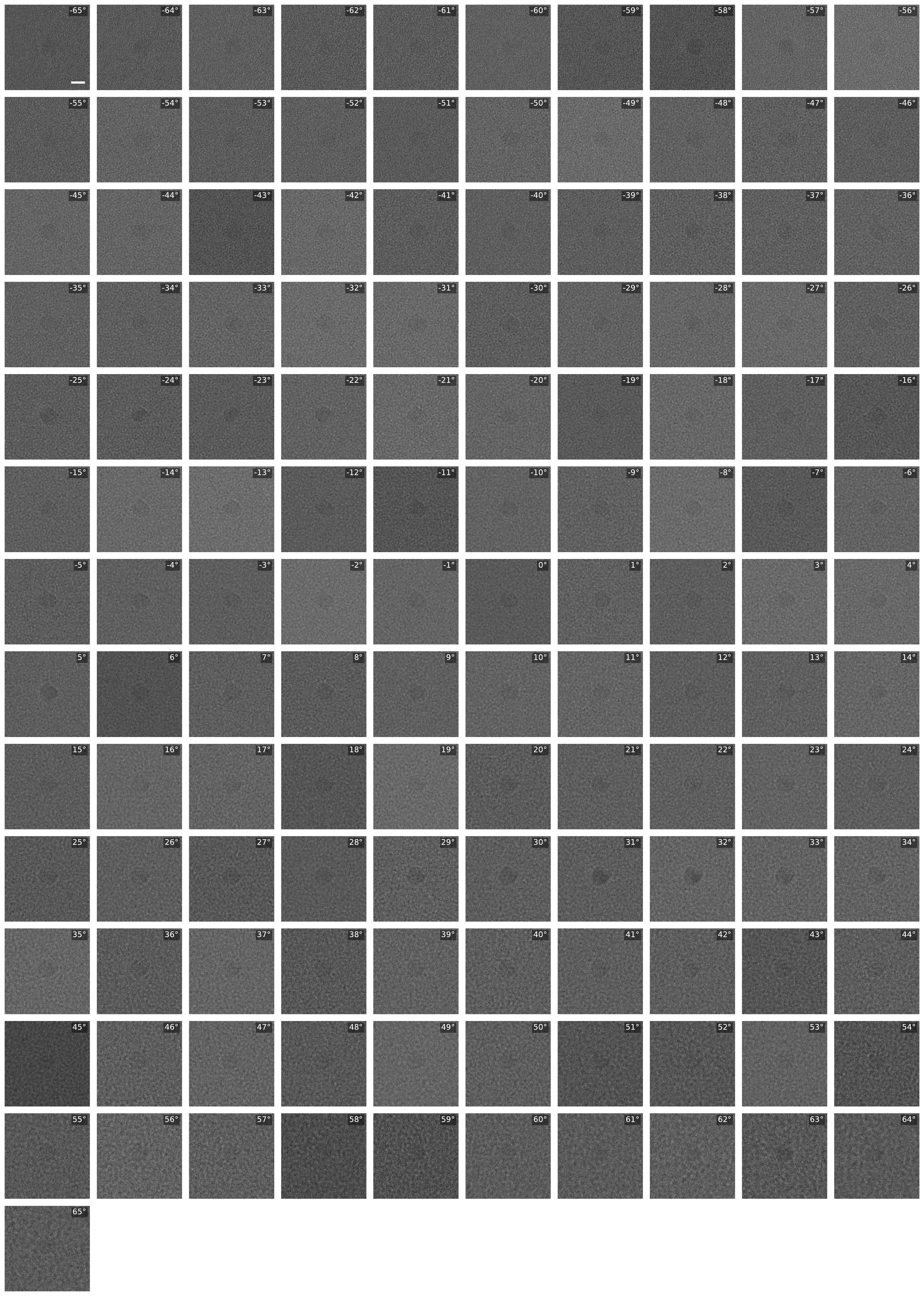}
\caption{
    \textbf{Experimental HRTEM tomographic tilt-series of a Co$_3$O$_4$ nanoparticle (second defocus).} 
    Each image represents a motion-corrected sum of frames. Image intensities are represented in grayscale. Scale bar: 5 nm.
}
\end{figure}

\begin{figure}[H]\label{fig:Co3O4_tilt_series2}
\centering
\includegraphics[width=0.9\textwidth]{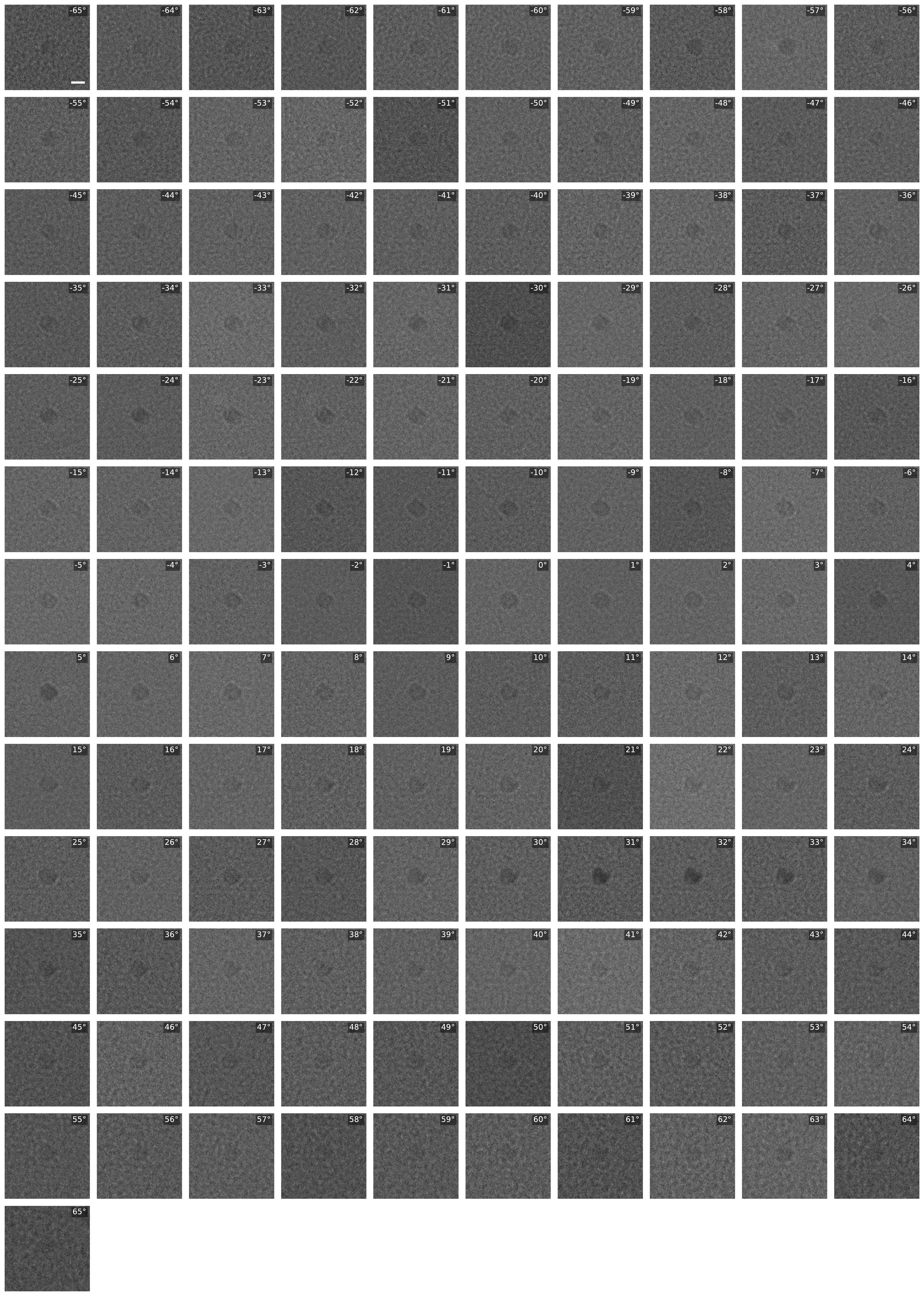}
\caption{
    \textbf{Experimental HRTEM tomographic tilt-series of a Co$_3$O$_4$ nanoparticle (third defocus).} 
    Each image represents a motion-corrected sum of frames. Image intensities are represented in grayscale. Scale bar: 5 nm.
}
\end{figure}

\begin{figure}[H]\label{fig:positivity}
\centering
\includegraphics[width=1\textwidth]{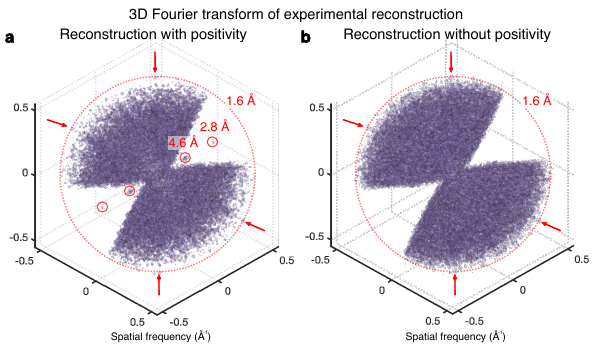}
\caption{
    \textbf{Comparison of missing wedge recovery with and without positivity constraint}
    \textbf{a,} 3D Fourier transform of the experimental PhaseT3M reconstruction of the Co$_3$O$_4$ nanoparticle shown in \hyperref[fig:exp_recon]{Fig. 2b}, displaying recovered diffraction peaks within the missing wedge due to the positivity constraint.
    \textbf{b,} 3D Fourier transform of the reconstruction without the positivity constraint, showing no recovery of diffraction peaks in the missing wedge.
}
\end{figure}

\begin{figure}[H]\label{fig:multi_single}
\centering
\includegraphics[width=1\textwidth]{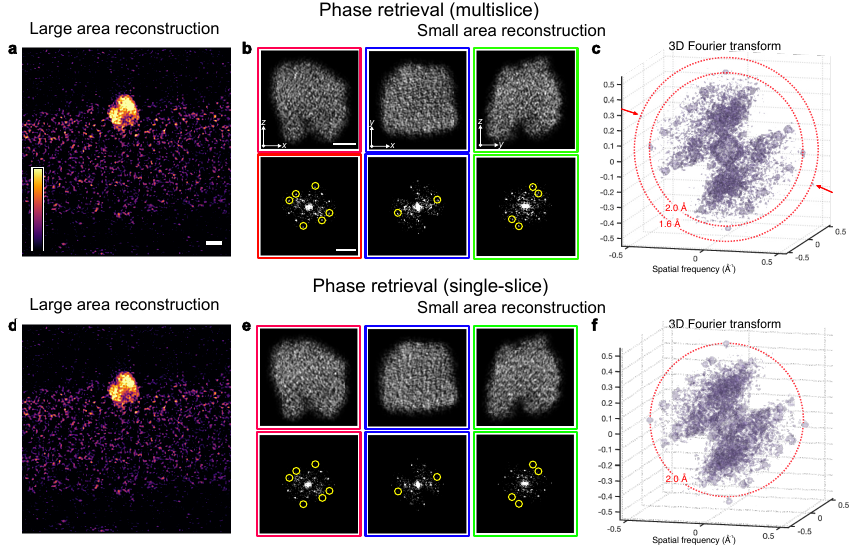}
\caption{
    \textbf{Comparison of multislice and single-slice PhaseT3M reconstructions.} 
    \textbf{a, d,} 4~\AA-thick near-central 3D slices of the Co$_3$O$_4$ nanoparticle and its carbon support, reconstructed using the multislice (a) and single-slice (d) phase retrieval methods. Scale bar: 4 nm.
    \textbf{b, e,} 2~\AA-thick slices extracted from the 3D reconstructions using multislice (b) and single-slice (e) phase retrieval methods. Slice colors correspond to the slice locations indicated in \hyperref[fig:exp_recon]{Fig. 2b–d}. Yellow circles highlight diffraction peaks that are present only in the multislice reconstruction and absent (or very faint) in the single-slice result. Image intensities are represented in grayscale, and Fourier transform intensities are shown on a logarithmic scale in grayscale. Top panel scale bar, 1 nm; bottom panel scale bar, 0.4~\AA$^{-1}$.
    \textbf{c, f,} 3D Fourier transforms of the reconstructions using the multislice (c) and single-slice (f) phase retrieval methods. Panel (c) is the same as shown in \hyperref[fig:exp_recon]{Fig. 2e}. The 3D Fourier transform of the multislice reconstruction (c) shows a maximum resolution of 1.6~\AA, whereas that of the single-slice reconstruction (f) shows a maximum resolution of 2.0~\AA.
}
\end{figure}

\begin{figure}[H]\label{fig:error_curve_Co3O4}
\centering
\includegraphics[width=0.7\textwidth]{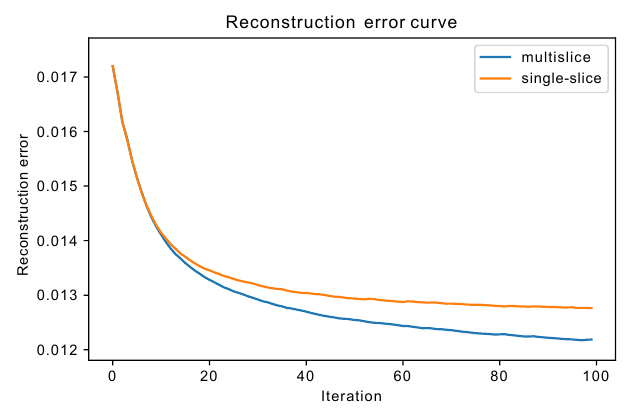}
\caption{
    \textbf{Reconstruction error curves of the Co$_3$O$_4$ nanoparticle.} 
    Reconstruction error curves are shown for both multislice reconstruction (corresponding to Fig. 2 and Supplementary Fig. 5a-c) and single-slice reconstruction (Supplementary Fig. 5d-f). In all cases, the reconstruction errors decrease and converge, confirming the stability of the iterative process. Additionally, the multislice reconstruction shows consistently lower error compared to the single-slice approach, indicating that incorporating a multislice model to account for multiple scattering leads to improved reconstruction accuracy. Source data for line plots in this figure are provided as a Source Data file.
}
\end{figure}

\begin{figure}[H]\label{fig:defocus_variation}
\centering
\includegraphics[width=1\textwidth]{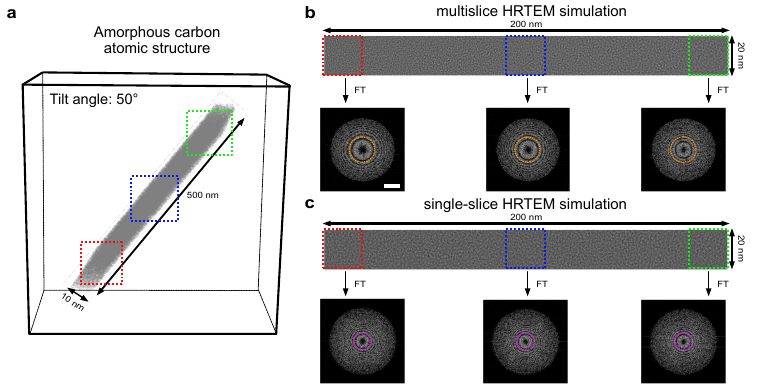}
\caption{
    \textbf{HRTEM simulations for a tilted sample.} 
    \textbf{a,} Atomic structure of an amorphous carbon sample tilted at 50 degrees. The atomic structure was generated by randomly placing atoms in a box of dimensions 20 $\times$ 500 $\times$ 10 nm$^3$, then tilted by 50 degrees. The tilted atomic structure is placed in a box of dimensions 20 $\times$ 500 $\times$ 500 nm$^3$ for input volume of HRTEM simulation.
    \textbf{b-c,} HRTEM simulation results of the tilted sample using the multislice method (b) and the single-slice method (c), showing simulated HRTEM images (field of view: 200 $\times$ 20 nm$^2$) and their corresponding Fourier transforms of cropped regions. The red, blue, and green boxes indicate different image locations. Orange and purple rings mark the first Thon ring for the Fourier transform of the red-colored region, serving as a visual guideline. Simulation parameters included a pixel size of 1 nm and a slice thickness of 4 nm (used only for the multislice method). Fourier transform intensities are shown on a logarithmic scale in grayscale. Scale bar: 2 nm$^{-1}$.
}
\end{figure}

\begin{figure}[H]\label{fig:res_elec_dose}
\centering
\includegraphics[width=1\textwidth]{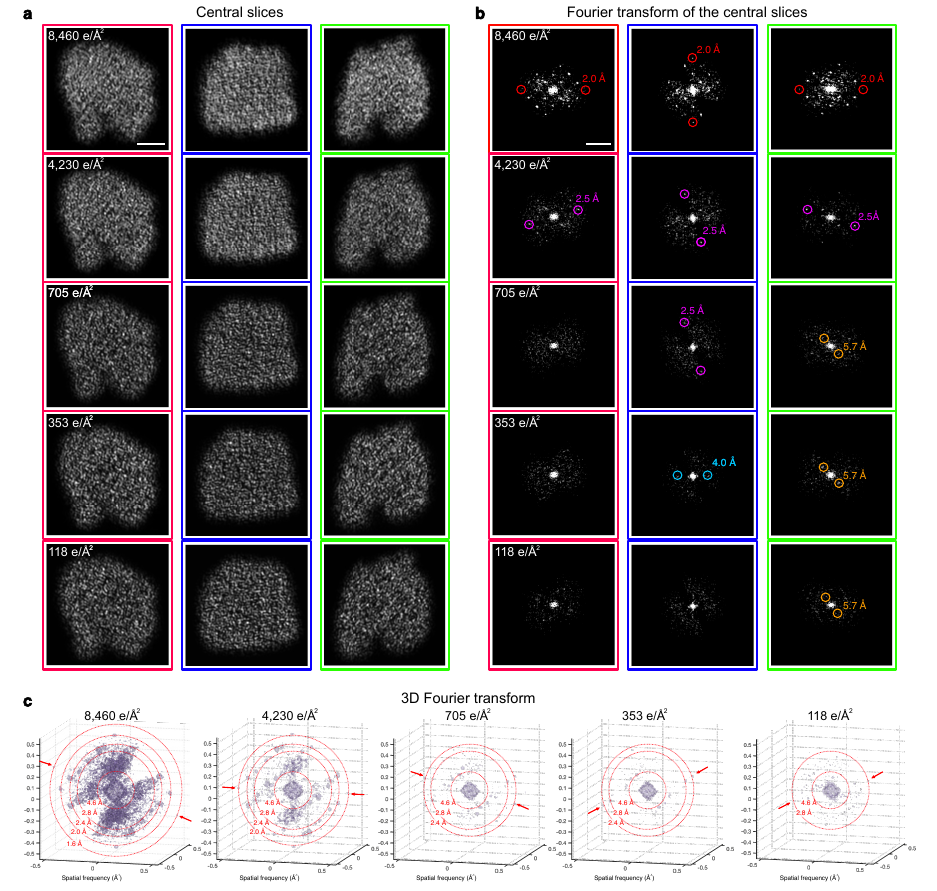}
\caption{
    \textbf{Phase retrieval reconstructions of the Co$_3$O$_4$ nanoparticle at varying electron dose.} 
    \textbf{a,} Central 2 Å-thick slices from the reconstructed volumes, shown along three orthogonal directions for different electron dose levels: 8460~$e^{-}$/\AA$^2$ (using all 24 subframes), 4230~$e^{-}$/\AA$^2$ (12 subframes), 705~$e^{-}$/\AA$^2$ (2 subframes), 353~$e^{-}$/\AA$^2$ (1 subframe), and 118~$e^{-}$/\AA$^2$ (using only one defocus image and one subframe). Image intensities are represented in grayscale. Scale bar: 2 nm. 
    \textbf{b,} 2D Fourier transforms of the corresponding central slices. Fourier transform intensities are shown on a logarithmic scale in grayscale. Scale bar: 0.4~$e^{-}$/\AA$^2$. Colored rings indicate diffraction peaks corresponding to resolution limits. Colored frames indicate slice orientations using the same convention as in \hyperref[fig:exp_recon]{Fig. 2}.
    \textbf{c,} 3D Fourier transforms of the reconstructed volumes at different electron dose levels. The red dotted ring marks a reference resolution shell, and red arrows indicate visible 3D diffraction peaks.
}
\end{figure}

\begin{figure}[H]\label{fig:simul_Co3O4_dose_effect}
\centering
\includegraphics[width=1\textwidth]{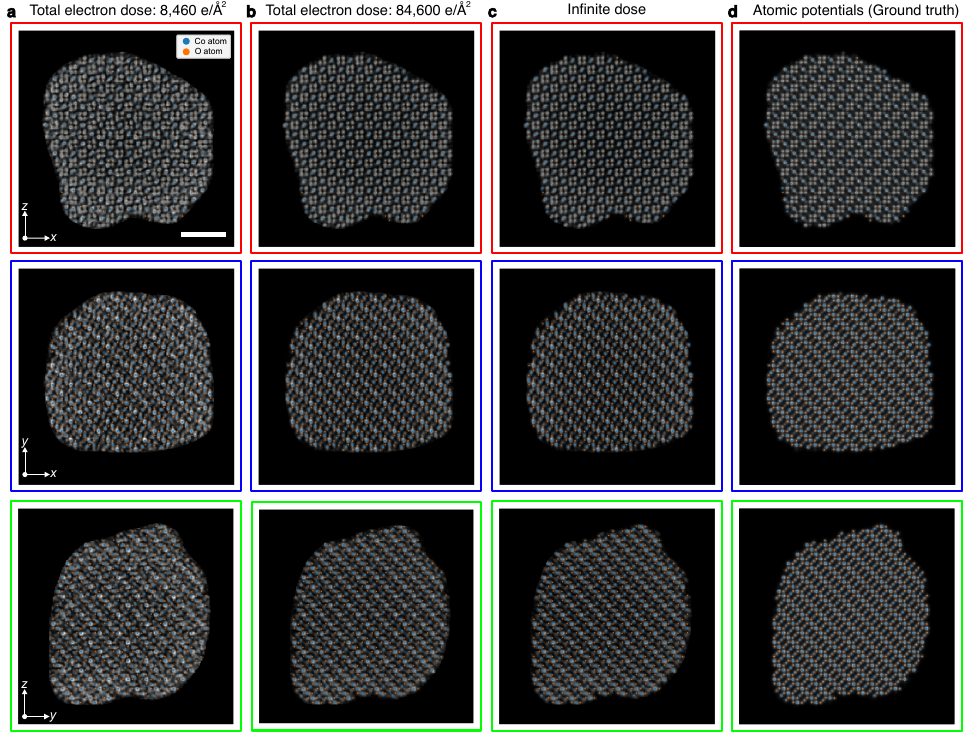}
\caption{
    \textbf{Simulated phase retrieval reconstruction varying electron dose and atomic potential of a Co$_3$O$_4$ nanoparticle.} 
    \textbf{a-c,} Central 2 Å-thick slices from the masked simulated reconstructed volume obtained with a total electron dose of 8,460~$e^{-}$/\AA$^2$ (a), 84,600~$e^{-}$/\AA$^2$ (b), and infinite (c), shown along three orthogonal directions.
    \textbf{d,} Central 2 Å-thick slices from 3D atomic potential of a Co$_3$O$_4$ nanoparticle, visualized along three orthogonal directions.
    All reconstructions were performed under the same experimental conditions: the experimental defocus values and C$_3$ value, optimized pixel size of 0.52 Å, and a tilt range from -65° to 65° with 1° angular steps. We also applied a low-pass filter of 0.8~\AA$^{-1}$, consistent with that used in the experimental reconstruction. For the dose-effect test, we used a clean model of a Co$_3$O$_4$ nanoparticle without alignment error and carbon background. The atomic model of the Co$_3$O$_4$ nanoparticle is overlaid with reconstructed intensities in panels (a-d). Colored frames indicate slice orientations, following the same convention as in Fig. 3. Image intensities are represented in grayscale. Scale bar: 1 nm.
}
\end{figure}

\begin{figure}[H]\label{fig:large_area_rec}
\centering
\includegraphics[width=1\textwidth]{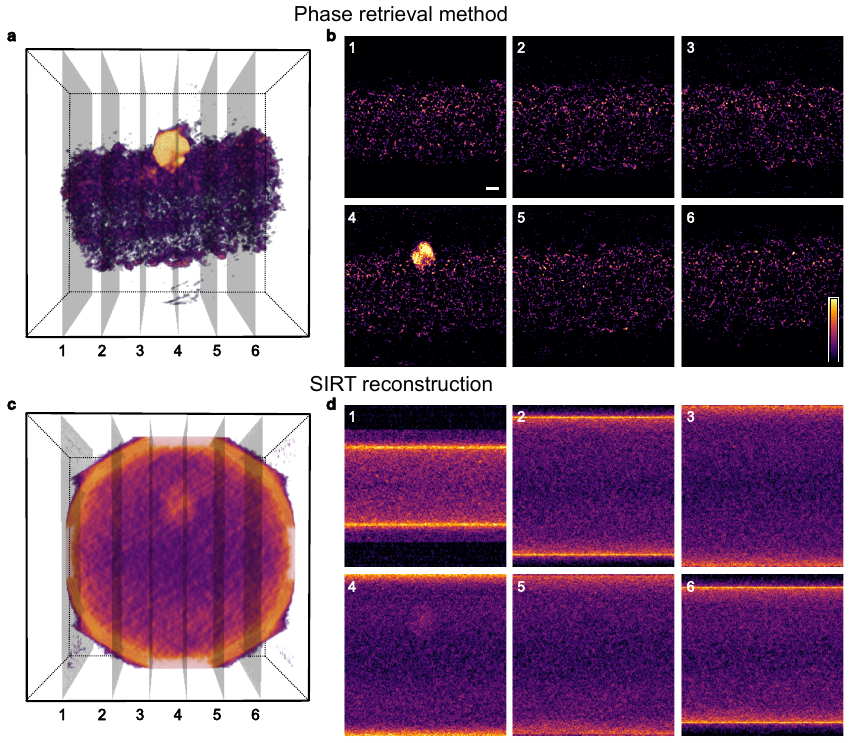}
\caption{
    \textbf{Slices from large-area PhaseT3M reconstructions for the Co$_3$O$_4$ nanoparticle.} 
    \textbf{a, c,} 3D density maps of the Co$_3$O$_4$ nanoparticle and the carbon support reconstructed using phase retrieval (a) and SIRT (c). The 3D volumes were reconstructed with a voxel size of 2.08~\AA.
    \textbf{b, d,} 4~\AA-thick slices extracted from the 3D reconstructions in (a) and (c), respectively. The numbers in (a) and (c) indicate the slice positions corresponding to the numbered slices shown here. All color scales represent intensity in the inferno colormap. Scale bar: 4 nm.
}
\end{figure}

\begin{figure}[H]\label{fig:small_area_rec}
\centering
\includegraphics[width=1\textwidth]{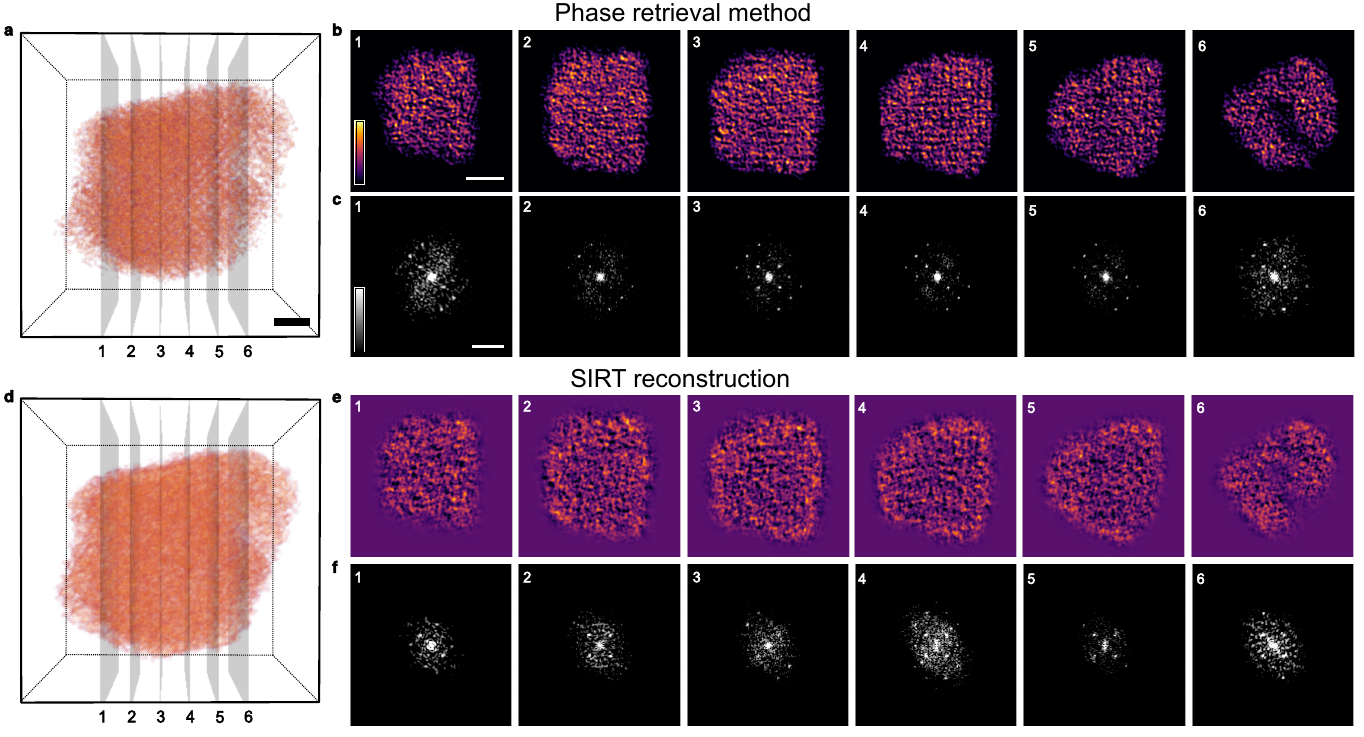}
\caption{
    \textbf{Slices from small-area 3D PhaseT3M Reconstructions of the Co$_3$O$_4$ nanoparticle and their 2D Fourier transforms.}
    \textbf{a, d,} 3D density maps of the Co$_3$O$_4$ nanoparticle reconstructed using phase retrieval (a) and SIRT (f), after applying a 3D mask to remove the carbon support intensity.    
    The volumes were reconstructed with a voxel size of 0.52~\AA. Scale bar: 1 nm.
    \textbf{b, e,} 1~\AA-thick slices extracted from the 3D reconstructions shown in (a) and (d), respectively. The slice numbers in (b) and (e) correspond to the positions indicated in the 3D reconstructions in (a) and (d), respectively. The color scale represents intensity in the inferno colormap. Scale bar: 2 nm
    \textbf{c, f,} 2D Fourier transforms of the slices in (b) and (e), respectively. Fourier transform intensities are shown on a logarithmic scale in grayscale. Scale bar: 0.4 \AA$^{-1}$. 
}
\end{figure}

\begin{figure}[H]\label{fig:conven_tomo_res}
\centering
\includegraphics[width=1\textwidth]{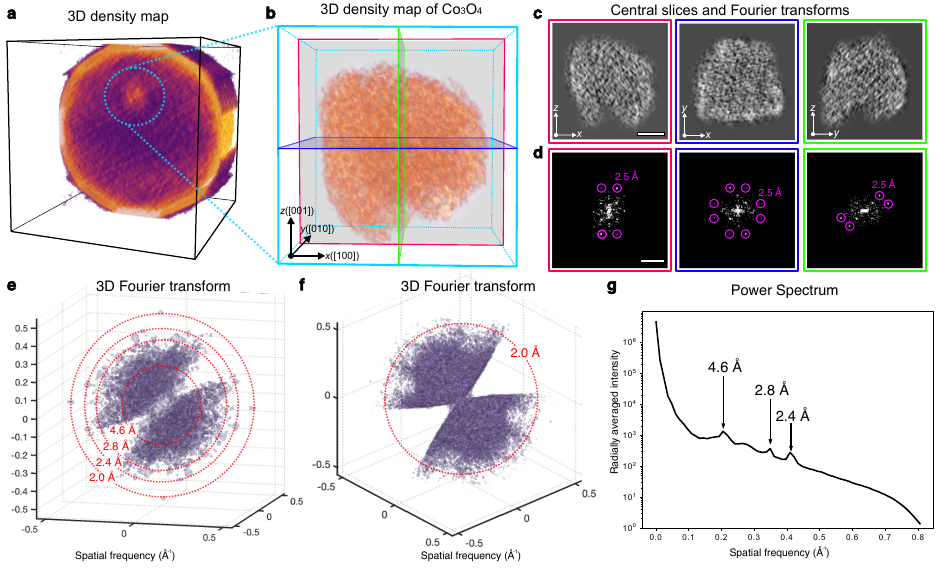}
\caption{
    \textbf{Conventional SIRT reconstruction and resolution analysis.} 
    \textbf{a,} Half-sectioned 3D density map illustrating the internal cross-section of the full 3D SIRT reconstruction of a Co$_3$O$_4$ nanoparticle embedded in a carbon support. 
    \textbf{b,} 3D density map of the Co$_3$O$_4$ nanoparticle after applying a 3D mask to remove the carbon support. Scale bar: 1 nm.
    \textbf{c,} 2-\AA~thick central slices of the 3D reconstruction in (b), with each color frame corresponding to the slice color in the 3D reconstruction. Image intensities are represented in grayscale. Scale bar: 2 nm.
    \textbf{d,} 2D Fourier transform of the central slice shown in the top panel. Fourier transform intensities are shown on a logarithmic scale in grayscale. Scale bar: 0.4 \AA$^{-1}$.
    \textbf{e,} 3D Fourier transform of the reconstruction in (b), displaying diffraction peaks at 2.0 \AA\ resolution.
    \textbf{f,} 3D Fourier transform of the reconstruction in (b) from a different view, highlighting the missing wedge region. The image clearly shows no diffraction peaks within the missing wedge.
    \textbf{g,} Power spectrum of the 3D reconstruction in (b), indicating a resolution of 2.4 \AA.
    Source data for line plots in this figure are provided as a Source Data file.
}
\end{figure}

\begin{figure}[H]\label{fig:ctf}
\centering
\includegraphics[width=0.7\textwidth]{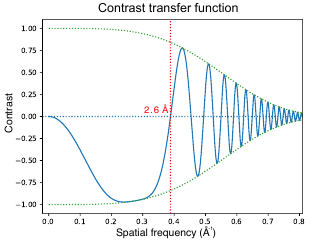}
\caption{
    \textbf{Contrast transfer function (CTF) curve under the experimental conditions.}
    The Scherzer defocus value was calculated using an optimized spherical aberration (C$_3$) value of 2.3 mm. The green dotted line represents the chromatic envelope function, computed using typical chromatic aberration (C$_\mathrm{C}$) of 2.7 mm and an energy spread ($\Delta E$) of 0.7 eV for the Titan Krios G3i microscope. The resulting contrast transfer function (CTF) indicates a diffraction-limited resolution of 2.6 \AA.
    Source data for line plots in this figure are provided as a Source Data file.
}
\end{figure}

\begin{figure}[H]\label{fig:SSNR}
\centering
\includegraphics[width=1\textwidth]{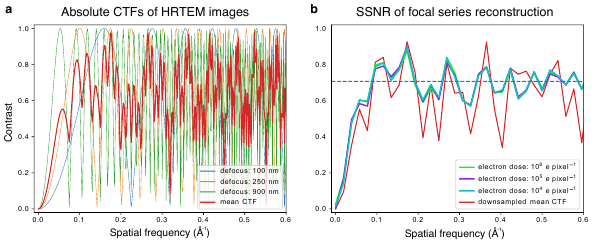}
\caption{
    \textbf{Spectral Signal-to-Noise Ratio (SSNR) of 2D Focal series reconstruction.} 
    \textbf{a,} Simulated CTFs for defocus values of 100, 250, and 900 nm.
    \textbf{b,} SSNRs of focal series reconstructions using HRTEM images with defocus values of 100, 250, and 900 nm. For simplicity, the SSNRs of the focal series reconstruction are simulated under idealized conditions, assuming no chromatic aberration and using a single-slice reconstruction model.
    Source data for line plots in this figure are provided as a Source Data file.
}
\end{figure}

\begin{figure}[H]\label{fig:HIV_tilt_series}
\centering
\includegraphics[width=1\textwidth]{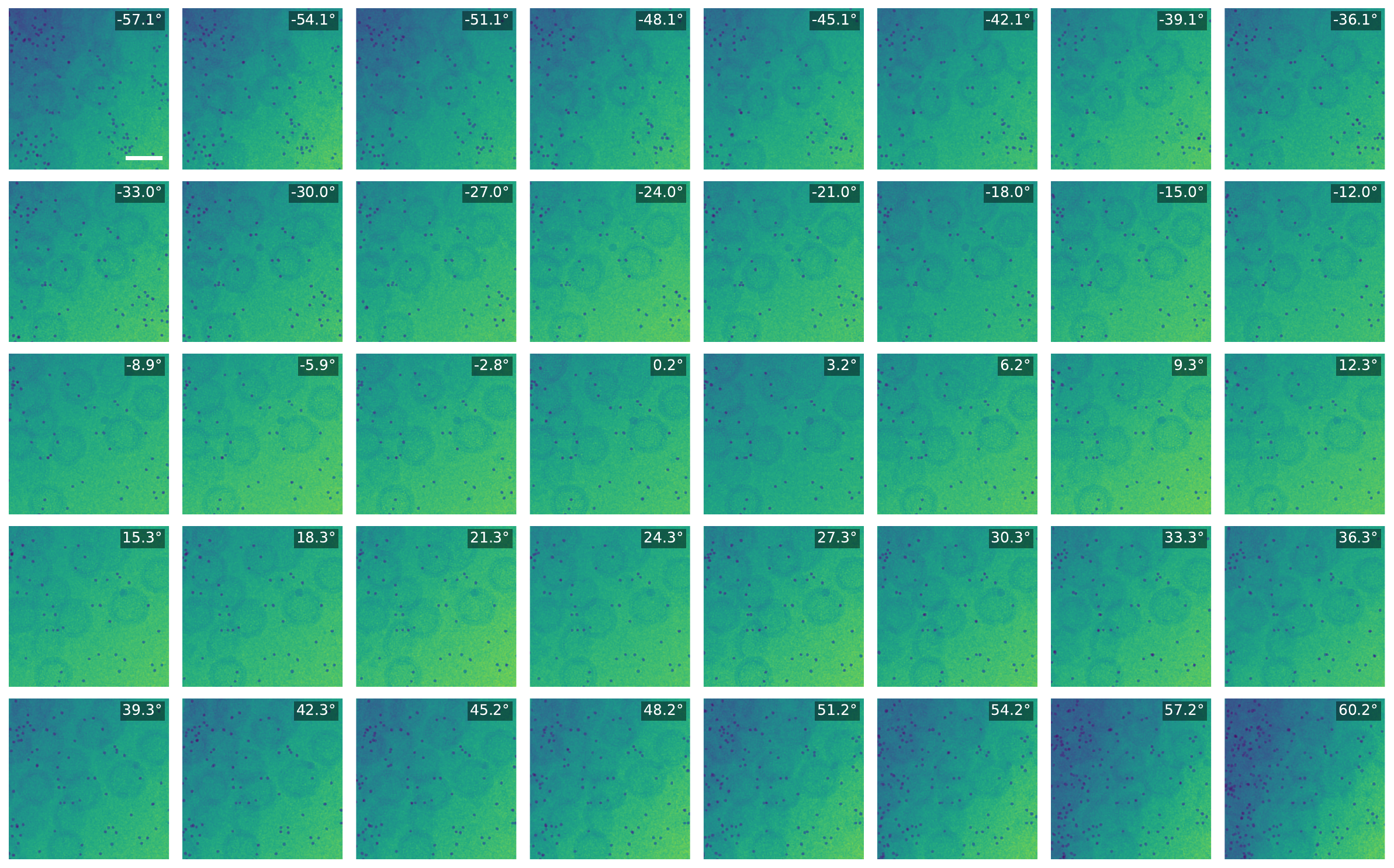}
\caption{
    \textbf{Experimental HRTEM tomographic tilt-series of an HIV-1 particle (EMPIAR-10164).} 
    Each image represents a motion-corrected sum of frames. Image intensities are represented in the viridis colormap. Scale bar: 100 nm.
}
\end{figure}

\begin{figure}[H]\label{fig:hiv_projs}
\centering
\includegraphics[width=1\textwidth]{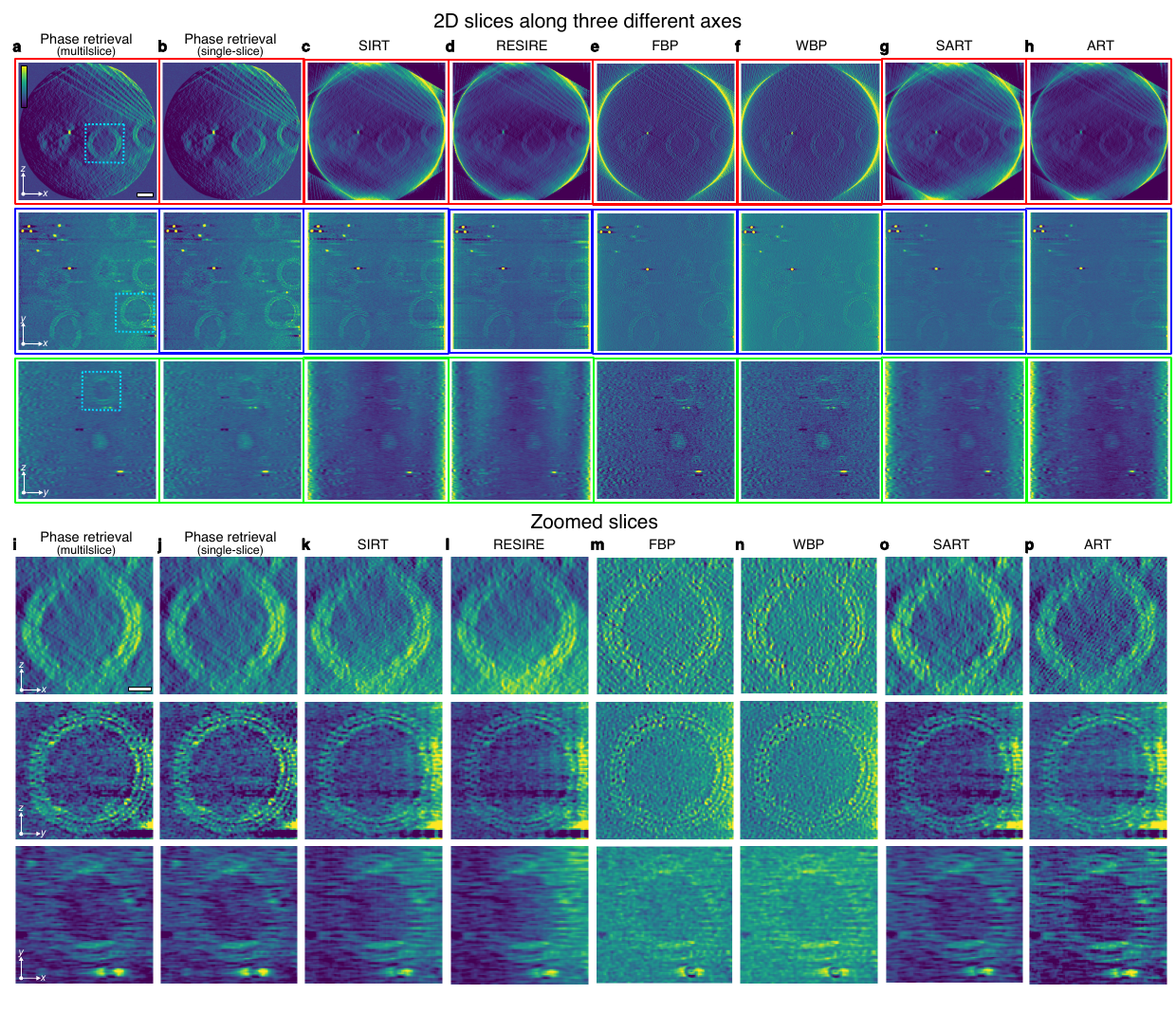}
\caption{
    \textbf{Slices of HIV-1 reconstructions using different methods.}
    \textbf{a-h.} 12-nm thick slices extracted from the reconstruction volumes are shown for multislice PhaseT3M (a), single-slice PhaseT3M (b), SIRT (c), RESIRE (d), FBP (e), WBP (f), SART (g), and ART (h). The colored frames correspond to the slice positions shown in \hyperref[fig:bio_tomo]{Fig. 5a–b}.
    \textbf{i-p.} Enlarged views of the boxed regions are displayed for multislice PhaseT3M (i), single-slice PhaseT3M (j), SIRT (k), RESIRE (l), FBP (m), WBP (n), SART (o), and ART (p). Image intensities are represented in the viridis colormap. Scale bars: top panels, 50 nm; bottom panels, 10 nm.
}
\end{figure}

\begin{figure}[H]\label{fig:ramlak_filter}
\centering
\includegraphics[width=1\textwidth]{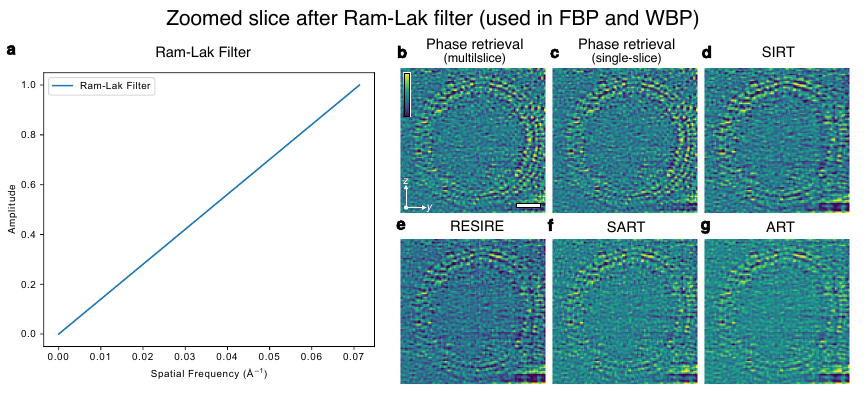}
\caption{
    \textbf{Ram-Lak filter plot and reconstruction results after applying the filter.}
    \textbf{a,} 1D Ram-Lak filter plot used in FBP and WBP. The filter suppresses low-frequency information, enhancing high-frequency components.
    \textbf{b–g,} Enlarged 12-nm thick slices extracted from the reconstruction volumes after the Ram-Lak filter for multislice PhaseT3M (b), single-slice PhaseT3M (c), SIRT (d), RESIRE (e), SART (f), and ART (g). The frames correspond to the slice positions shown in Fig. 5d–h and Supplementary Fig. 16i–p. Image intensities are represented in the viridis colormap. Scale bar: 10 nm.
    Source data for line plots in this figure are provided as a Source Data file.
}
\end{figure}

\begin{figure}[H]\label{fig:average_frc}
\centering
\includegraphics[width=0.75\textwidth]{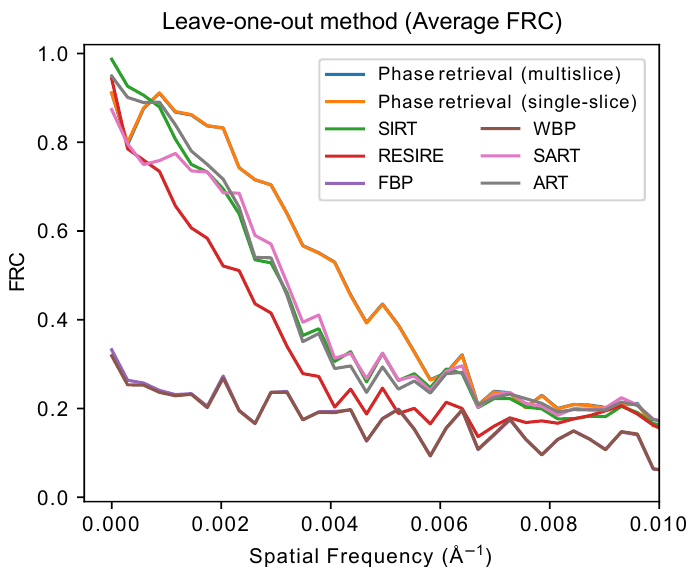}
\caption{
    \textbf{Leave-one-out method (Average Fourier ring correlation)}
    Average Fourier ring correlation (FRC) between each experimental tilt-angle HRTEM projection and the corresponding calculated projection generated from the 3D volume reconstructed using a leave-one-out strategy (excluding the respective tilt angle from the reconstruction). The FRC was first calculated for each tilt angle individually, and the results were then averaged across all tilt angles to produce the final curves. FRC curves are compared across different methods: multislice phase retrieval (PhaseT3M), single-slice phase retrieval (PhaseT3M), real space iterative reconstruction (RESIRE), simultaneous iterations reconstruction technique (SIRT), filtered back projection (FBP), weighted back projection (WBP), simultaneous algebraic reconstruction technique (SART), and algebraic reconstruction technique (ART).
    Source data for line plots in this figure are provided as a Source Data file.
}
\end{figure}

\begin{figure}[H]\label{fig:rfactor}
\centering
\includegraphics[width=0.75\textwidth]{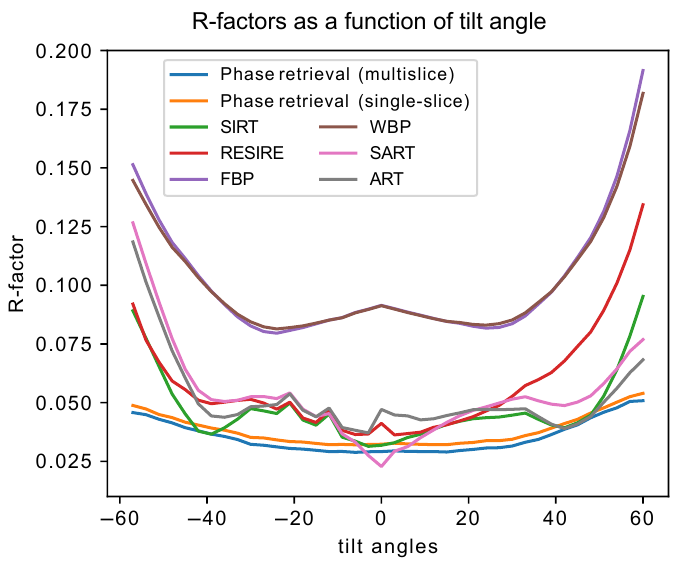}
\caption{
    \textbf{R-factors across different reconstruction methods.}
    R-factors between the measured and calculated projections are shown for each tilt angle, comparing different reconstruction methods. All PhaseT3M results, except for the SART result at the zero-degree tilt, demonstrate superior performance, indicating that the PhaseT3M projections closely match the experimental data. The average R-factors across all tilt angles are 0.035 for multislice PhaseT3M, 0.038 for single-slice PhaseT3M, 0.047 for SIRT, 0.058 for RESIRE, 0.101 for FBP, 0.101 for WBP, 0.053 for SART, and 0.052 for ART.
    Source data for line plots in this figure are provided as a Source Data file.
}
\end{figure}

\begin{figure}[H]\label{fig:error_curve_EMPIAR}
\centering
\includegraphics[width=0.75\textwidth]{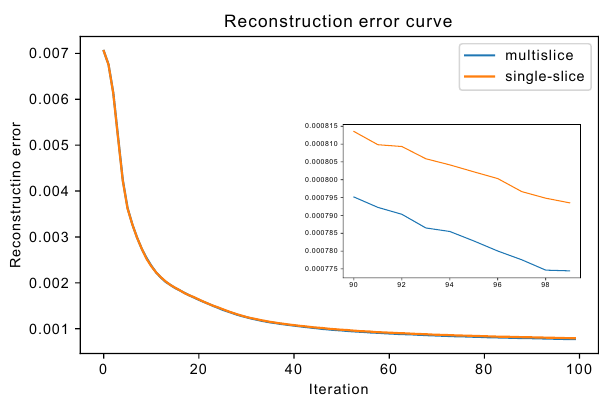}
\caption{
    \textbf{Reconstruction error curves of the HIV-1 particles.} 
    Reconstruction error curves are shown for both multislice reconstruction (Fig. 5 and Supplementary Fig. 16a) and single-slice reconstruction (Supplementary Fig. 16b). In all cases, the reconstruction errors decrease and converge, confirming the stability of the iterative process. An inset shows a zoomed-in view of the reconstruction errors from 90 to 100 iterations to highlight the differences. The multislice reconstruction errors are slightly lower than single-slice reconstruction errors, but they are nearly identical, unlike the Co$_3$O$_4$ nanoparticle case, due to the dominance of weak scattering from light elements.
    Source data for line plots in this figure are provided as a Source Data file.
}
\end{figure}

\subsection{Pseudo code for 3D reconstruction algorithm} \label{pseudocode}

\RestyleAlgo{ruled}
\begin{algorithm}[H]
\DontPrintSemicolon
\SetNoFillComment
\SetAlgoLined
    \caption{Psuedo code to reconstruct 3D electrostatic potential}
    
    \IncMargin{1.5em}
    
    \SetKwInOut{Input}{input}
    \SetKwInOut{Output}{output}
    
    \Input{A tilt and focal series of measured HRTEM images $\left\{ I_{\theta, \Delta f} \right\}$, 
    Tilt angle set $\{ \theta \}$, Defocus value set $\{ \Delta f \}$, \newline
    Step size of the gradient descent method $\alpha$, \newline
    Number of iterations: $N_\textrm{iter}$ }
   
    \Output{3D electrostatic potential $V\left(x,y,z \right)$}

    \BlankLine
    $V_{(1)}\left(x,y,z\right) \gets 0$ \;

    \BlankLine
    \tcc{main reconstruction}
    $\psi_{0}\left(x,y \right) \gets 1$ \; incident parallel electron wave function
    
    $P(q_x,q_y) \gets$ free-space propagation with thickness $\Delta z$ \;
    \BlankLine    
    \For{$i$ in \{1 to $N_\textrm{iter}$\}}{

        \For{$\theta$ in tilt angle set \{$\theta$\}}{

            $V_\textrm{rot}\left(x,y,z\right) = {R_\theta V}_{(i)}\left(x,y,z\right)$ 

            \BlankLine 
            \tcc{forward propagation}
            $\psi_{1}\left(x,y\right) \gets \psi_{0}$ initial function\;
            
            ${{V}_m^\textrm{2D}(x,y)} \gets$ projected potential set from $V_\textrm{rot}\left(x,y,z\right)$ \;
            
            $\left\{t_m(x,y)\right\} \gets$ transmission function set from \{${V}_m^\textrm{2D}\left(x,y\right)$\}  \;
            
            \For{m in \{1 to $N_{z}$\} }{
                $\psi_{m+1}\left(x,y\right) \gets \mathcal{F}^{-1}[P(q_x,q_y)\mathcal{F}[t_m\left(x,y\right)\psi_{m}\left(x,y\right)]]$ 
            }
            
            $\psi_{\textrm{exit}, \theta}\left(x,y\right) \gets {\psi}_{N_{z+1}}\left(x,y\right)$

            \For{$\Delta f$ in \{$\Delta f$\} }{
            $\psi_{\textrm{final}, \theta, \Delta f}\left(x,y\right) \gets \mathcal{F}^{-1}[\psi_{\textrm{exit}, \theta}\left(q_x,q_y\right) \exp{(-i \chi \left(q_x,q_y\right) )}]$ \;
            
            ${\hat{I}}_{\theta, \Delta f}(x,y) \gets  {|\mathcal{F}{(\psi}_{\textrm{final}, \theta, \Delta f}\left(x,y\right))|}^2  $
            }

            \BlankLine 
            \BlankLine  
            \tcc{backpropagation}
            $\psi_{N_{z+1}}^{\textrm{back}}(q_x,q_y) \gets  \sum_{\Delta f} \exp\left(i \chi{(\Delta f)}\right) \mathcal{F} \left( \psi_{\text{final}, \theta, \Delta f} - \sqrt{I_{\theta, \Delta f}}  \frac{\psi_{\text{final}, \theta, \Delta f}}{\left| \psi_{\text{final}, \theta, \Delta f} \right|}  \right)$ \; 

            \For{m in \{$N_{z}$ to 1\} }{
                $\psi_{m}^{\textrm{back}}\left(x,y\right)\gets {\mathcal{F}^{-1}[P}_m^\ast\left(q_x,q_y\right)\psi_{m+1}^{\textrm{back}}(q_x,q_y)]$ \;
                
                $\mathrm{\nabla}_{V}\mathcal{E}^2\left(x,y,m\right) \gets \operatorname{Re} \left({-i\ t}_m^\ast\left(x,y\right)\psi_m^\ast{\left(x,y\right)\psi}_{m}^{\textrm{back}}\left(x,y\right) \right)$ \;
                
                $\psi_{m}^{\textrm{back}}(q_x,q_y) \gets \mathcal{F}[t_m^\ast\left(x,y\right)\ \psi_{m}^{\textrm{back}}\left(x,y\right)]$  \;
                
            }

            \BlankLine 
            \tcc{update 3D potential}
            $V_\textrm{rot}\left(x,y,z\right) \gets V_\textrm{rot}\left(x,y,z\right)-\alpha_{V}\ \mathrm{\nabla}_{V}\mathcal{E}^2(x,y,z) $ \;
            
            $V_{(i)}\left(x,y,z\right)\gets {R_\theta^{-1}V}_\textrm{rot}\left(x,y,z\right)$ \;
            
        }
    }
    $V\left(x,y,z\right) \gets V_{(N_\textrm{iter})}\left(x,y,z\right)$ 
    
\end{algorithm}

\end{document}